\documentclass[10pt]{amsart}
\usepackage{graphicx}
\usepackage{amsfonts,amssymb,amsmath}
\newcommand*{\abs}[1]{\left\lvert#1\right\rvert}   % absolute value of #1
\newcommand*{\set}[1]{\left\{#1\right\}}           % set of ...
\newcommand*{\brs}[1]{\left(#1\right)}             % brackets
\newcommand*{\norm}[1]{\left\Vert#1\right\Vert}    % norm of #1
\newcommand*{\hilb}{\mathcal H}                     % Hilbert space

\newcommand*{\mbC}{{\mathbb C}}
\newcommand*{\mbR}{{\mathbb R}}
\newcommand*{\mbZ}{{\mathbb Z}}

\newcommand*{\mbQ}{{\mathbb Q}}

\newcommand*{\Tr}{\operatorname{Tr}}        % usual trace
\renewcommand*{\det}{\operatorname{det}}     % determinant

    \newtheorem{thm*}{Theorem}

    \newtheorem{lemma*}{Lemma}    %for AMS \newtheorem*

    \newtheorem{rems*}{Remark}   %for AMS \newtheorem*

\newcommand{\ndef}{\newcommand*}
\def\rndef{\renewcommand}

\ndef{\myaddress}[1]{\begin{center} \it\small #1 \end{center}}

\ndef{\clA}{{\mathcal A}} \ndef{\rmA}{{\mathrm A}} \ndef{\mbA}{{\mathbb A}} \ndef{\bfA}{{\mathbf A}} \ndef{\euA}{{\EuScript A}} \ndef{\frA}{{\mathfrak A}}
\ndef{\clB}{{\mathcal B}} \ndef{\rmB}{{\mathrm B}} \ndef{\mbB}{{\mathbb B}} \ndef{\bfB}{{\mathbf B}} \ndef{\euB}{{\EuScript B}} \ndef{\frB}{{\mathfrak B}}
\ndef{\clC}{{\mathcal C}} \ndef{\rmC}{{\mathrm C}}                          \ndef{\bfC}{{\mathbf C}} \ndef{\euC}{{\EuScript C}} \ndef{\frC}{{\mathfrak C}}
\ndef{\clD}{{\mathcal D}} \ndef{\rmD}{{\mathrm D}} \ndef{\mbD}{{\mathbb D}} \ndef{\bfD}{{\mathbf D}} \ndef{\euD}{{\EuScript D}} \ndef{\frD}{{\mathfrak D}}
\ndef{\clE}{{\mathcal E}} \ndef{\rmE}{{\mathrm E}} \ndef{\mbE}{{\mathbb E}} \ndef{\bfE}{{\mathbf E}} \ndef{\euE}{{\EuScript E}} \ndef{\frE}{{\mathfrak E}}
\ndef{\clF}{{\mathcal F}} \ndef{\rmF}{{\mathrm F}} \ndef{\mbF}{{\mathbb F}} \ndef{\bfF}{{\mathbf F}} \ndef{\euF}{{\EuScript F}} \ndef{\frF}{{\mathfrak F}}
\ndef{\clG}{{\mathcal G}} \ndef{\rmG}{{\mathrm G}} \ndef{\mbG}{{\mathbb G}} \ndef{\bfG}{{\mathbf G}} \ndef{\euG}{{\EuScript G}} \ndef{\frG}{{\mathfrak G}}
\ndef{\clH}{{\mathcal H}} \ndef{\rmH}{{\mathrm H}} \ndef{\mbH}{{\mathbb H}} \ndef{\bfH}{{\mathbf H}} \ndef{\euH}{{\EuScript H}} \ndef{\frH}{{\mathfrak H}}
\ndef{\clI}{{\mathcal I}} \ndef{\rmI}{{\mathrm I}} \ndef{\mbI}{{\mathbb I}} \ndef{\bfI}{{\mathbf I}} \ndef{\euI}{{\EuScript I}} \ndef{\frI}{{\mathfrak I}}
\ndef{\clJ}{{\mathcal J}} \ndef{\rmJ}{{\mathrm J}} \ndef{\mbJ}{{\mathbb J}} \ndef{\bfJ}{{\mathbf J}} \ndef{\euJ}{{\EuScript J}} \ndef{\frJ}{{\mathfrak J}}
\ndef{\clK}{{\mathcal K}} \ndef{\rmK}{{\mathrm K}} \ndef{\mbK}{{\mathbb K}} \ndef{\bfK}{{\mathbf K}} \ndef{\euK}{{\EuScript K}} \ndef{\frK}{{\mathfrak K}}
\ndef{\clL}{{\mathcal L}} \ndef{\rmL}{{\mathrm L}} \ndef{\mbL}{{\mathbb L}} \ndef{\bfL}{{\mathbf L}} \ndef{\euL}{{\EuScript L}} \ndef{\frL}{{\mathfrak L}}
\ndef{\clM}{{\mathcal M}} \ndef{\rmM}{{\mathrm M}} \ndef{\mbM}{{\mathbb M}} \ndef{\bfM}{{\mathbf M}} \ndef{\euM}{{\EuScript M}} \ndef{\frM}{{\mathfrak M}}
\ndef{\clN}{{\mathcal N}} \ndef{\rmN}{{\mathrm N}}                          \ndef{\bfN}{{\mathbf N}} \ndef{\euN}{{\EuScript N}} \ndef{\frN}{{\mathfrak N}}
\ndef{\clO}{{\mathcal O}} \ndef{\rmO}{{\mathrm O}} \ndef{\mbO}{{\mathbb O}} \ndef{\bfO}{{\mathbf O}} \ndef{\euO}{{\EuScript O}} \ndef{\frO}{{\mathfrak O}}
\ndef{\clP}{{\mathcal P}} \ndef{\rmP}{{\mathrm P}} \ndef{\mbP}{{\mathbb P}} \ndef{\bfP}{{\mathbf P}} \ndef{\euP}{{\EuScript P}} \ndef{\frP}{{\mathfrak P}}
\ndef{\clQ}{{\mathcal Q}} \ndef{\rmQ}{{\mathrm Q}}                          \ndef{\bfQ}{{\mathbf Q}} \ndef{\euQ}{{\EuScript Q}} \ndef{\frQ}{{\mathfrak Q}}
\ndef{\clR}{{\mathcal R}} \ndef{\rmR}{{\mathrm R}}                          \ndef{\bfR}{{\mathbf R}} \ndef{\euR}{{\EuScript R}} \ndef{\frR}{{\mathfrak R}}
\ndef{\clS}{{\mathcal S}} \ndef{\rmS}{{\mathrm S}} \ndef{\mbS}{{\mathbb S}} \ndef{\bfS}{{\mathbf S}} \ndef{\euS}{{\EuScript S}} \ndef{\frS}{{\mathfrak S}}
\ndef{\clT}{{\mathcal T}} \ndef{\rmT}{{\mathrm T}} \ndef{\mbT}{{\mathbb T}} \ndef{\bfT}{{\mathbf T}} \ndef{\euT}{{\EuScript T}} \ndef{\frT}{{\mathfrak T}}
\ndef{\clU}{{\mathcal U}} \ndef{\rmU}{{\mathrm U}} \ndef{\mbU}{{\mathbb U}} \ndef{\bfU}{{\mathbf U}} \ndef{\euU}{{\EuScript U}} \ndef{\frU}{{\mathfrak U}}
\ndef{\clV}{{\mathcal V}} \ndef{\rmV}{{\mathrm V}} \ndef{\mbV}{{\mathbb V}} \ndef{\bfV}{{\mathbf V}} \ndef{\euV}{{\EuScript V}} \ndef{\frV}{{\mathfrak V}}
\ndef{\clW}{{\mathcal W}} \ndef{\rmW}{{\mathrm W}} \ndef{\mbW}{{\mathbb W}} \ndef{\bfW}{{\mathbf W}} \ndef{\euW}{{\EuScript W}} \ndef{\frW}{{\mathfrak W}}
\ndef{\clX}{{\mathcal X}} \ndef{\rmX}{{\mathrm X}} \ndef{\mbX}{{\mathbb X}} \ndef{\bfX}{{\mathbf X}} \ndef{\euX}{{\EuScript X}} \ndef{\frX}{{\mathfrak X}}
\ndef{\clY}{{\mathcal Y}} \ndef{\rmY}{{\mathrm Y}} \ndef{\mbY}{{\mathbb Y}} \ndef{\bfY}{{\mathbf Y}} \ndef{\euY}{{\EuScript Y}} \ndef{\frY}{{\mathfrak Y}}
\ndef{\clZ}{{\mathcal Z}} \ndef{\rmZ}{{\mathrm Z}}                          \ndef{\bfZ}{{\mathbf Z}} \ndef{\euZ}{{\EuScript Z}} \ndef{\frZ}{{\mathfrak Z}}

\ndef{\tA}{{\widetilde A}} \ndef{\tcA}{{\widetilde\clA}} \ndef{\ttcA}{\widetilde{\tcA}} \ndef{\sfA}{{\textsf A}} \ndef{\ttA}{\widetilde{\tA}} \ndef{\dzA}{{A^\sharp}}
\ndef{\tB}{{\widetilde B}} \ndef{\tcB}{{\widetilde\clB}} \ndef{\ttcB}{\widetilde{\tcB}} \ndef{\sfB}{{\textsf B}} \ndef{\ttB}{\widetilde{\tB}} \ndef{\dzB}{{B^\sharp}}
\ndef{\tC}{{\widetilde C}} \ndef{\tcC}{{\widetilde\clC}} \ndef{\ttcC}{\widetilde{\tcC}} \ndef{\sfC}{{\textsf C}} \ndef{\ttC}{\widetilde{\tC}} \ndef{\dzC}{{C^\sharp}}
\ndef{\tD}{{\widetilde D}} \ndef{\tcD}{{\widetilde\clD}} \ndef{\ttcD}{\widetilde{\tcD}} \ndef{\sfD}{{\textsf D}} \ndef{\ttD}{\widetilde{\tD}} \ndef{\dzD}{{D^\sharp}}
\ndef{\tE}{{\widetilde E}} \ndef{\tcE}{{\widetilde\clE}} \ndef{\ttcE}{\widetilde{\tcE}} \ndef{\sfE}{{\textsf E}} \ndef{\ttE}{\widetilde{\tE}} \ndef{\dzE}{{E^\sharp}}
\ndef{\tF}{{\widetilde F}} \ndef{\tcF}{{\widetilde\clF}} \ndef{\ttcF}{\widetilde{\tcF}} \ndef{\sfF}{{\textsf F}} \ndef{\ttF}{\widetilde{\tF}} \ndef{\dzF}{{F^\sharp}}
\ndef{\tG}{{\widetilde G}} \ndef{\tcG}{{\widetilde\clG}} \ndef{\ttcG}{\widetilde{\tcG}} \ndef{\sfG}{{\textsf G}} \ndef{\ttG}{\widetilde{\tG}} \ndef{\dzG}{{G^\sharp}}
\ndef{\tH}{{\widetilde H}} \ndef{\tcH}{{\widetilde\clH}} \ndef{\ttcH}{\widetilde{\tcH}} \ndef{\sfH}{{\textsf H}} \ndef{\ttH}{\widetilde{\tH}} \ndef{\dzH}{{H^\sharp}}
\ndef{\tI}{{\widetilde I}} \ndef{\tcI}{{\widetilde\clI}} \ndef{\ttcI}{\widetilde{\tcI}} \ndef{\sfI}{{\textsf I}} \ndef{\ttI}{\widetilde{\tI}} \ndef{\dzI}{{I^\sharp}}
\ndef{\tJ}{{\widetilde J}} \ndef{\tcJ}{{\widetilde\clJ}} \ndef{\ttcJ}{\widetilde{\tcJ}} \ndef{\sfJ}{{\textsf J}} \ndef{\ttJ}{\widetilde{\tJ}} \ndef{\dzJ}{{J^\sharp}}
\ndef{\tK}{{\widetilde K}} \ndef{\tcK}{{\widetilde\clK}} \ndef{\ttcK}{\widetilde{\tcK}} \ndef{\sfK}{{\textsf K}} \ndef{\ttK}{\widetilde{\tK}} \ndef{\dzK}{{K^\sharp}}
\ndef{\tL}{{\widetilde L}} \ndef{\tcL}{{\widetilde\clL}} \ndef{\ttcL}{\widetilde{\tcL}} \ndef{\sfL}{{\textsf L}} \ndef{\ttL}{\widetilde{\tL}} \ndef{\dzL}{{L^\sharp}}
\ndef{\tM}{{\widetilde M}} \ndef{\tcM}{{\widetilde\clM}} \ndef{\ttcM}{\widetilde{\tcM}} \ndef{\sfM}{{\textsf M}} \ndef{\ttM}{\widetilde{\tM}} \ndef{\dzM}{{M^\sharp}}
\ndef{\tN}{{\widetilde N}} \ndef{\tcN}{{\widetilde\clN}} \ndef{\ttcN}{\widetilde{\tcN}} \ndef{\sfN}{{\textsf N}} \ndef{\ttN}{\widetilde{\tN}} \ndef{\dzN}{{N^\sharp}}
\ndef{\tO}{{\widetilde O}} \ndef{\tcO}{{\widetilde\clO}} \ndef{\ttcO}{\widetilde{\tcO}} \ndef{\sfO}{{\textsf O}} \ndef{\ttO}{\widetilde{\tO}} \ndef{\dzO}{{O^\sharp}}
\ndef{\tP}{{\widetilde P}} \ndef{\tcP}{{\widetilde\clP}} \ndef{\ttcP}{\widetilde{\tcP}} \ndef{\sfP}{{\textsf P}} \ndef{\ttP}{\widetilde{\tP}} \ndef{\dzP}{{P^\sharp}}
\ndef{\tQ}{{\widetilde Q}} \ndef{\tcQ}{{\widetilde\clQ}} \ndef{\ttcQ}{\widetilde{\tcQ}} \ndef{\sfQ}{{\textsf Q}} \ndef{\ttQ}{\widetilde{\tQ}} \ndef{\dzQ}{{Q^\sharp}}
\ndef{\tR}{{\widetilde R}} \ndef{\tcR}{{\widetilde\clR}} \ndef{\ttcR}{\widetilde{\tcR}} \ndef{\sfR}{{\textsf R}} \ndef{\ttR}{\widetilde{\tR}} \ndef{\dzR}{{R^\sharp}}
\ndef{\tS}{{\widetilde S}} \ndef{\tcS}{{\widetilde\clS}} \ndef{\ttcS}{\widetilde{\tcS}} \ndef{\sfS}{{\textsf S}} \ndef{\ttS}{\widetilde{\tS}} \ndef{\dzS}{{S^\sharp}}
\ndef{\tT}{{\widetilde T}} \ndef{\tcT}{{\widetilde\clT}} \ndef{\ttcT}{\widetilde{\tcT}} \ndef{\sfT}{{\textsf T}} \ndef{\ttT}{\widetilde{\tT}} \ndef{\dzT}{{T^\sharp}}
\ndef{\tU}{{\widetilde U}} \ndef{\tcU}{{\widetilde\clU}} \ndef{\ttcU}{\widetilde{\tcU}} \ndef{\sfU}{{\textsf U}} \ndef{\ttU}{\widetilde{\tU}} \ndef{\dzU}{{U^\sharp}}
\ndef{\tV}{{\widetilde V}} \ndef{\tcV}{{\widetilde\clV}} \ndef{\ttcV}{\widetilde{\tcV}} \ndef{\sfV}{{\textsf V}} \ndef{\ttV}{\widetilde{\tV}} \ndef{\dzV}{{V^\sharp}}
\ndef{\tW}{{\widetilde W}} \ndef{\tcW}{{\widetilde\clW}} \ndef{\ttcW}{\widetilde{\tcW}} \ndef{\sfW}{{\textsf W}} \ndef{\ttW}{\widetilde{\tW}} \ndef{\dzW}{{W^\sharp}}
\ndef{\tX}{{\widetilde X}} \ndef{\tcX}{{\widetilde\clX}} \ndef{\ttcX}{\widetilde{\tcX}} \ndef{\sfX}{{\textsf X}} \ndef{\ttX}{\widetilde{\tX}} \ndef{\dzX}{{X^\sharp}}
\ndef{\tY}{{\widetilde Y}} \ndef{\tcY}{{\widetilde\clY}} \ndef{\ttcY}{\widetilde{\tcY}} \ndef{\sfY}{{\textsf Y}} \ndef{\ttY}{\widetilde{\tY}} \ndef{\dzY}{{Y^\sharp}}
\ndef{\tZ}{{\widetilde Z}} \ndef{\tcZ}{{\widetilde\clZ}} \ndef{\ttcZ}{\widetilde{\tcZ}} \ndef{\sfZ}{{\textsf Z}} \ndef{\ttZ}{\widetilde{\tZ}} \ndef{\dzZ}{{Z^\sharp}}

\ndef{\bfa}{{\mathbf a}}
\ndef{\bfb}{{\mathbf b}}
\ndef{\bfc}{{\mathbf c}}
\ndef{\bfd}{{\mathbf d}}

\ndef{\euu}{{\EuScript u}}

  \ndef{\eps}{\varepsilon}
\let\geq\geqslant

\ndef{\lims}[1]{\lim\limits_{#1}}
\ndef{\sums}[1]{\sum\limits_{#1}}
\ndef{\ints}[1]{\int_{#1}}
\ndef{\sups}[1]{\sup\limits_{#1}}
\ndef{\liminfty}[1]{\lims{#1\to\infty}}
\ndef{\suminf}[1]{\sums{#1=1}^\infty}

\ndef{\limo}[1]{\omega\mbox{-}\!\!\!\lims{#1\to\infty}}          % \omega limit
\ndef{\limL}[1]{\rmL\mbox{-}\!\!\!\lims{#1\to\infty}}            % ``L" limit
\ndef{\limLOne}[1]{\clL_1\mbox{-}\!\lims{#1}}
\ndef{\tildelimo}[1]{\tilde\omega\mbox{-}\!\!\!\lims{#1\to\infty}}
\ndef{\slim}{\mathrm{s}\mbox{-}\!\!\lim}          % strong limit
\ndef{\wlim}{\mathrm{w}\mbox{-}\!\lim}          % strong limit

\ndef{\Aut}{\operatorname{Aut}}      % group of automorphisms
\ndef{\Ch}{\operatorname{ch}}        % Chern character
\ndef{\End}{\operatorname{End}}      % group of endomorphisms
\ndef{\Hom}{\operatorname{Hom}}      % group of homomorphisms
\rndef{\ker}{\operatorname{ker}}      % kernel of an operator
\ndef{\coker}{\operatorname{coker}}      % kernel of an operator
\ndef{\im}{\operatorname{im}}        % image of an operator
\ndef{\Log}{\operatorname{Log}}      % logarithm
\ndef{\OP}{\operatorname{OP}}        % abstract PDO's
\ndef{\Op}{\operatorname{Op}}        % abstract PDO's
\ndef{\Symb}{\operatorname{Symb}}    % symbol
\ndef{\Wres}{\operatorname{Wres}}    % Wodzicki residue
\ndef{\cl}{\operatorname{cl}}        % Clifford
\ndef{\com}{\operatorname{com}}
\ndef{\const}{\operatorname{const}}  % constant
\ndef{\conv}{\operatorname{conv}}    % convex hull
\ndef{\Var}{\operatorname{Var}}
\ndef{\Cov}{\operatorname{Cov}}

\ndef{\detFK}[1]{\Delta\brs{#1}} % Fuglede-Kadison's determinant
\ndef{\detFKrel}[2]{\Delta_{#2}\brs{#1}} % Fuglede-Kadison's determinant

\ndef{\adj}{\operatorname{adj}}    % diagonal operator
\ndef{\diag}{\operatorname{diag}}    % diagonal operator
\ndef{\dist}{\operatorname{dist}}    % distance
\ndef{\dom}{\operatorname{dom}}      % domain
\ndef{\ec}{\operatorname{ec}}        % essential codimension
\ndef{\id}{\mathrm{Id}}                        % identity operator
\ndef{\ind}{\operatorname{ind}}      % index
\ndef{\mydeg}{\operatorname{deg}}    % degree of a diff. form
\ndef{\op}{\operatorname{op}}
\ndef{\rank}{\operatorname{rank}}
\ndef{\res}{\operatorname{res}}      % residue
\ndef{\ran}{\operatorname{ran}}      % range
\ndef{\sflow}{\operatorname{sf}}     % spectral flow
\ndef{\isf}{\operatorname{isf}}      % infinitesimal spectral flow
\ndef{\sign}{\operatorname{sign}}    % signum (a la C.-Ph.)
\ndef{\sgn}{\operatorname{sgn}}      % signum (a la Connes)
\ndef{\sing}{\operatorname{sing}}    % singular
\ndef{\supp}{\operatorname{supp}}    % support
\ndef{\tr}{\operatorname{tr}}        % trace
\ndef{\var}{\operatorname{var}}      % variation of measure
\ndef{\vol}{\operatorname{vol}}      % volume or volume form
\ndef{\wn}{\operatorname{wn}}        % winding number
\ndef{\wres}{\operatorname{wres}}    % Wodzicki residue
\rndef{\Im}{\operatorname{Im}}       % imaginary part of an operator
\rndef{\Re}{\operatorname{Re}}       % real part of an operator

\ndef{\prng}[1]{\mathrm R_{#1}} % {[\rng {#1}]}          % projection onto the range of #1
\ndef{\pker}[1]{\mathrm N_{#1}} % {[\ker {#1}]}          % projection onto the kernel of #1
\ndef{\rprng}[2]{\mathrm R_{#1}^{#2}}           % projection onto the relative range of #1
\ndef{\rpker}[2]{\mathrm N_{#1}^{#2}}           % projection onto the relative kernel of #1
\ndef{\rsupp}[1]{\supp_r(#1)}
\ndef{\lsupp}[1]{\supp_l(#1)}
\ndef{\rslv}[1]{R_z(#1)}      % resolvent
\ndef{\HH}{H}                 % initial operator of perturbation theory
\ndef{\tHH}{\tilde \HH}       % final operator of perturbation theory
\ndef{\VV}{V}                 % perturbation operator of perturbation theory
\ndef{\Rz}{R_z}               % resolvent of the initial operator
\ndef{\tRz}{\tR_z}            % resolvent of the final operator
\ndef{\psif}[1]{#1^{[1]}} % {\psi_{#1}}  % divided difference
\ndef{\WPlus}[1]{W_{#1}(\mbR)} %\ndef{\CPlus}[1]{C^{#1+}(\mbR)}
\ndef{\bndl}{\xi}                         % vector bundle
\ndef{\bndlA}{\eta}                       % vector bundle
\ndef{\GlueMap}{\varphi}                  % glue map of a bundle
\ndef{\ChartMap}{h}                       % chart diffeomorphism map of a manifold
\ndef{\chern}{\ensuremath{\mathrm{ch}}}
\ndef{\hilba}{\clH^{(a)}}                    % absolutely continuous part of Hilbert space (wrt to a s.a. operator)
\ndef{\hilbs}{\clH^{(s)}}                    % singular part of Hilbert space (wrt to a s.a. operator)
   \ndef{\hilbasargument}{(\hilb)} %{(\hilb)}
\ndef{\LpH}[1]{\clL_{#1}\hilbasargument}       % the set of ...
\ndef{\saLpH}[1]{\clL_{sa}^{#1}\hilbasargument}       % the set of ...
\ndef{\clBH}{\clB\hilbasargument}              % the set of BOUNDED linear operators on Hilbert space
\ndef{\ubBH}{\clB_1\hilbasargument}            % the unit ball of the algebra of all BOUNDED linear operators on Hilbert space
\ndef{\clCH}{\clC\hilbasargument}              % the set of CLOSED DENSELY-DEFINED linear operators on Hilbert space
\ndef{\clKH}{\clK\hilbasargument}              % the set of COMPACT operators
\ndef{\clFH}{\clF\hilbasargument}              % the set of BOUNDED FREDHOLM operators
\ndef{\clUH}{\clU\hilbasargument}              % the set of UNITARIES on Hilbert space
\ndef{\clCFH}{{\clC\clF}\hilbasargument}       % the set of CLOSED DENSELY-DEFINED FREDHOLM OPERATORS on Hilbert space
\ndef{\saBH}{\clB_{sa}\hilbasargument}         % the set of S.-A. BOUNDED operators on Hilbert space
\ndef{\saCH}{\clC_{sa}\hilbasargument}         % the set of CLOSED DENSELY-DEFINED S.-A. operators on Hilbert space
\ndef{\saFH}{\clF_{sa}\hilbasargument}         % the set of BOUNDED FREDHOLM s.-a. operators
\ndef{\saKH}{\clK_{sa}\hilbasargument}         % the set of COMPACT S.-A. operators
\ndef{\saCFH}{\clC\clF_{sa}\hilbasargument}    % the set of CLOSED DENSELY-DEFINED S.-A. FREDHOLM operators on Hilbert space
\ndef{\clUFH}{\clU\clF\hilbasargument}         % the set of UNITARIES such that U+I is FREDHOLM
\ndef{\Uinj}{\clU_{inj}\hilbasargument}        % the set of UNITARIES such that U-I is injective
\ndef{\UFinj}{\clU\clF_{inj}\hilbasargument}   % the set of UNITARIES such that U-I is INJECTIVE and U+I is FREDHOLM

\ndef{\spproj}[2]{E^{#1}_{#2}}                      % spectral projection of #1
\ndef{\spprojb}[2]{E^{#2}_{#1}}                     % spectral projection of #1

\ndef{\LpN}[1]{\clL^{#1}(\clN,\tau)}     % noncommutative \mathcal L_p space
\ndef{\saLpN}[1]{\clL^{#1}_{sa}(\clN,\tau)} % s.-a. part of noncommutative \mathcal L_p space
\ndef{\rLpN}[1]{L^{#1}(\clN,\tau)}       % noncommutative L_p space
\ndef{\clAND}{(\clA,\clN,D)}             % spectral triple (A,N,D)
\ndef{\clBA}{{\clB(\clA)}}
\ndef{\saKN}{{\clK_{sa}(\clN,\tau)}}          % s.-a. \tau-compact operators
\ndef{\clKN}{{\clK(\clN,\tau)}}          % \tau-compact operators
\ndef{\clKtN}{{\clK(\tilde\clN,\tau)}}   % \tau-compact (maybe unbounded) operators
\ndef{\clFN}{{\clF(\clN,\tau)}}          % \tau-Fredholm operators
\ndef{\saFN}{{\clF_{sa}(\clN,\tau)}}     % self-adjoint \tau-Fredholm operators
\ndef{\clPN}{\clP(\clN)}                 % projections of N
\ndef{\clQN}{\clQ(\clN,\tau)}            % Calkin algebra N/K
\ndef{\infPN}{{\clP_\tau^\infty(\clN)}}  % infinite projections of N
\ndef{\clOF}[2]{\clF_{#1\mbox{-}#2}(\clN,\tau)}         % relatively Fredholm operators
\ndef{\oind}[2]{{\rm \tau\mbox{-}ind}_{#1\mbox{-}#2}}   % relative index
\ndef{\tind}{\tau\mbox{-}\ind}                  % semifinite index
\ndef{\DInd}{\ind_{\clD,\tau}}           % semifinite index
\ndef{\BF}{Breuer-Fredholm}              % Breuer-Fredholm
\ndef{\skewfred}[2]{$(#1\cdot #2)$ $\tau$\tire Fredholm}   % skew corner Fredholm
\ndef{\affl}{\eta}                       % affiliated
\ndef{\vNa}{von Neumann algebra}         % von Neumann algebra
\ndef{\nsf}{faithful normal semifinite } % normal semifinite faithful
\ndef{\taubrs}[1]{\tau\brackets{#1}}     % n.s.f. trace with brackets
\ndef{\sqbrs}[1]{[#1]}        % brackets
\ndef{\Sqbrs}[1]{\big[#1\big]}        % brackets
\ndef{\SqBrs}[1]{\Big[#1\Big]}        % brackets
\ndef{\domd}{\bigcap\limits_{n\ge 0} \dom\;\delta^n}         % domain of \delta^n's
\ndef{\DiffOP}{{\rm \clD}}
\ndef{\ADA}{\clA \cup [\clD,\clA]}
\ndef{\DixIdeal}[1]{\LpH{#1,\infty}}               % Dixmier ideal
\ndef{\dixideal}{\ell^{1,\infty}}                  % commutative Dixmier ideal
\ndef{\WDixIdeal}{\LpH{1,\mathrm w}}               % weak Dixmier ideal
\ndef{\DixIdealPos}[1]{\DixIdeal{#1}_+}            % positive part of Dixmier ideal
\ndef{\DixIdealN}[1]{\LpN{#1,\infty}}              % semifinite Dixmier ideal
\ndef{\DixIdealNPar}[2]{\clL^{#1,\infty}_{#2}(\clN,\tau)}    % semifinite Dixmier ideal
\ndef{\DixIdealNPos}[1]{\LpN{#1,\infty}_+}                   % positive part of semifinite Dixmier ideal
\ndef{\TrD}{\Tr_\omega}                                      % Dixmier trace
\ndef{\tauD}{{\tau_\omega}}                                  % semifinite Dixmier trace
\ndef{\ILogN}{\frac 1{\log(1+N)}}
\ndef{\DixNorm}[1]{\norm{#1}_{(1,\infty)}}                   % Dixmier norm
\ndef{\DixInt}[1]{\ints 0^t \mu_s(#1)\,ds}
\ndef{\DixIntL}[1]{\ints 0^{\lambda_{1/t}(#1)}\mu_s(#1)\,ds}
    \ndef{\SmallIdeal}{{\clL^{1, \mathrm w}}}
    \ndef{\SmallIdealMeas}{{\clL^{1, \mathrm w}_m}}
    \ndef{\DixIntII}[1]{\int_0^t \mu_s(#1)\,ds}
    \ndef{\DixIntf}[1]{\Phi_t(#1)}
    \ndef{\DixIntg}[1]{\Psi_t(#1)}

\ndef{\lpi}{\clL^{1,\pi}(\clN,\tau)}
\ndef{\strl}[1]{\stackrel \longrightarrow {#1}}
\ndef{\IIinfty}{$\mathrm{II}_\infty$\ }

\ndef{\fourier}[1]{\clF(#1)}          % Fourier transform of #1
\ndef{\HaarMeasBohrs}{\nu}            % Haar measure of the Bohr compact
\ndef{\BrownsMeas}{\mu}               % Brown's measure
\ndef{\BohrCont}[1]{\tilde{#1}}       % continuation of a function to the Bohr compact
\ndef{\APMean}{{M}}                   % mean value of a.p. function
\ndef{\CDSS}{{\clA_B}}                % Coburn-Douglas-Schaeffer-Singer's factor
\ndef{\matr}{{\rm Mat}}               % standard matrix algebra
\ndef{\seque}[1]{\ensuremath{\{#1_n\}_{n=1}^\infty}}    % sequence of numbers  a_1, a_2, ...
\ndef{\sequen}[2]{\ensuremath{\{#1_#2\}_{#2=1}^\infty}}    % sequence of numbers  a_1, a_2, ...
\ndef{\Seque}[1]{\ensuremath{\left(#1_0,#1_1,#1_2,\dots\right)}}    % sequence of numbers  a_1, a_2, ...
\ndef{\Cesaro}{H}                           % the Cesaro operator (on sequences)
\ndef{\CesaroRPlus}{M}                      % the Cesaro operator on positive semiaxis
\ndef{\Dilation}{D}                         % the dilation operator (on sequences)
\ndef{\Shift}{T}                            % the shift operator (on sequences)

\ndef{\TrNorm}[1]{\norm{#1}_1}              % trace norm of #1
\ndef{\HSNorm}[1]{\norm{#1}_2}              % Hilbert-Schmidt norm of #1
\ndef{\InftyNorm}[1]{\norm{#1}_\infty}      % uniform norm of #1
\ndef{\normQN}[1]{\norm{#1}_{\clQN}}        % Calkin norm of #1
\ndef{\clLpnorm}[2]{\norm{#2}_{\clL^{#1}}}    % $1,\infty$- trace norm of #1
\ndef{\clLnorm}[1]{\clLpnorm{1}{#1}}    % $1,\infty$- trace norm of #1

\ndef{\ccurve}{\gamma}                      % a curve in complex plane for Cauchy integral

\ndef{\Brs}[1]{\big(#1\big)}                % brackets
\ndef{\BRS}[1]{\Big(#1\Big)}                % brackets
\ndef{\scal}[2]{\left\langle #1,#2\right\rangle}               % scalar product
\ndef{\Scal}[1]{\left\langle #1\right\rangle}               % scalar product
\ndef{\precprec}{\prec\!\!\!\prec}
\ndef{\qeq}{\stackrel?=}
\ndef{\spectrum}[1]{\sigma_{#1}} %{\mathrm{Spec}(#1)}       % spectrum of an operator
\ndef{\spectruma}[1]{\sigma^{(a)}_{#1}} %{\mathrm{Spec}(#1)}       % absolutely continuous spectrum of an operator
\rndef{\emptyset}{\varnothing}                              % empty set
\ndef{\csupp}{c}                           % subscript for compactly supported functions
\ndef{\closure}[1]{\overline{#1}}
\ndef{\linspan}[1]{\mathrm{span}\,{#1}}
\ndef{\bddborel}[1]{B(#1)}                 % the space of bounded Borel functions on the measure space #1
\ndef{\charfunc}{\chi}
\ndef{\FrDer}{\euD}                        % Fr\'echet derivative
\ndef{\LieDer}[1]{\pounds_{#1}\,}          % Lie derivative
\ndef{\dds}{\left.\frac d{ds} \right|_{s = 0}}
\ndef{\ortcmp}[1]{#1^{\scriptscriptstyle \perp}}            % orthogonal complement of projection #1
\ndef{\Laplace}{\Delta}                    % Laplace operator

\ndef{\matrPQ}[3]
{
    \left(
      \begin{array}{cc}
        #1_{11} & #1_{12} \\
        #1_{21} & #1_{22}
      \end{array}
    \right)_{[#2,#3]}
}
\ndef{\margOK}{\marginpar{\bf \small OK}}

\newcounter{margcomcount}
\ndef{\margcom}[1]{\marginpar{\bf \small #1} \addtocounter{margcomcount}{1}
   \index{\indexcom{{\bf COMMENT: #1}}}}

\ndef{\mytimes}{\!\times\!}
\ndef{\sss}[1]{\subsubsection{}\label{#1}}

\rndef{\phi}{\varphi} \ndef{\OpenUnitDisk}{D}
\ndef{\RHS}{RHS}                            % right hand side
\ndef{\LHS}{LHS} %right and left hand side  % left hand side
\ndef{\ttt}{\Leftrightarrow}
\ndef{\then}{\Rightarrow}
\ndef{\tto}{\longrightarrow}
\ndef{\nno}{\nonumber\\}
\ndef{\newn}[1]{\index{#1} {\bfseries #1}}       % new notion
\ndef{\la}{\langle}
\ndef{\ra}{\rangle}
\ndef{\dbar}{{\;\bar{\phantom{o}} \!\!\!\! d}}
\ndef{\stl}[1]{\stackrel{\vbox to 0pt{\vss\hbox{$\scriptstyle #1$}}}}
\ndef{\mathcomment}[1]{{\hfill \qquad\qquad\qquad\text{by (#1)}}}        % for comments at the ends of lines with math formulas
\ndef{\mathcomm}[1]{{\hfill \qquad\qquad\qquad\qquad\qquad\text{#1}}}        % for comments at the ends of lines with math formulas
\ndef{\details}[1]{\smallskip\begin{center} {\bf Here:}
#1\end{center}\medskip} \ndef{\indexcom}[1]{ --- #1}
\ndef{\longsim}{\ \sim \ }              % for use in formulas.
\ndef{\tire}{-}              % for use in formulas.
\ndef{\intinfinf}{\int_{-\infty}^\infty}
     \ndef{\npartial}{\slash\!\!\!\partial}
     \ndef{\Heis}{\operatorname{Heis}}
     \ndef{\Solv}{\operatorname{Solv}}
     \ndef{\Spin}{\operatorname{Spin}}
     \ndef{\SO}{\operatorname{SO}}
     \ndef{\Index}{\operatorname{index}}
             \ndef{\p}{\partial}
             \ndef{\dd}{|\clD|}
             \ndef{\n}{\parallel}

\usepackage[matrix,arrow,curve]{xy}

\usepackage{tikz}
\usepackage{circuitikz}
\usepackage{listings}
\usepackage{color}
\newcommand{\writeanswer}[1]{}
\newcommand{\Spike}{{Spik\`e$\!\!\!$\'{}}\,}

\newcounter{nques}
\newcommand{\QuesA}[2]{\bigskip \addtocounter{nques}{1} %
   \noindent \mbox{{\bf\arabic{nques}.}\,[$#2$p]}}
\newcommand{\QuesB}[1]{\medskip}

\newcommand{\dash}{'}

\newcommand{\problemtype}[1]{}  %\margcom{\tiny \sf (#1)}}

\newcommand{\congruent}[3]{#1 \equiv #2 \ \ (\,\mathrm{mod} \ #3)}

\begin{document}
%\begin{center}
%   \includegraphics{1.mps}
%\end{center}

\title{MATLAB based language for generating \\ randomised multiple choice questions II}
\author{Nurulla Azamov}
\address{College of Science and Engineering
   \\ Flinders University 
   \\ South Rd, Tonsley, SA 5042 Australia}
\email{nurulla.azamov@flinders.edu.au}

%\keywords{resonance index, spectral flow}
% \subjclass[2000]{ %Mathematics Subject Classification (2000).
%    Primary 47A40;(???)
% }

\begin{abstract} In this work we present an improved version of a MATLAB based coding 
language for generating randomised multiple choice questions. This language has been 
successfully tested at Flinders University in quite a few mathematics topics.

In the second version of this language there are a number of improvements,
including 
\begin{enumerate}
  \item the so-called hash-tag operators which convert Matlab data structures into mathematics structures in LaTeX format,
  \item a possibility to keep spike-blocks in separate files and to use pointers to files,
  \item a few new types of questions, 
  \item the \verb!freeze! command which allows to freeze certain random variables 
        for the generation of alternative incorrect answers, 
  \item the text part \verb!\freeze! command which allows a certain part of a question's text
       not to be repeated in repeats of the question,
  \item numerous bugs have been fixed,
  \item and there are many new examples of questions generated by spike-blocks. 
\end{enumerate}

\vskip 1cm 
\begin{center}
  \small \sf \today, v\,8.5
\end{center}
\end{abstract}

\maketitle

\newpage

\tableofcontents

\vskip 1cm \section{Introduction}

This manual gives description of a special purpose 
language \Spike, which is 
designed to generate randomised multiple choice tests in mathematics
and other subjects. \Spike\ is written in and is based on MATLAB. 
The output which \Spike\ produces is a \LaTeX\ file with randomised individualised assignments.
This presumes that a potential user of \Spike\ is to be familiar
to a certain extent with both MATLAB and \LaTeX.

\Spike\ combines computing power of MATLAB with \LaTeX's typesetting power.
Anyone familiar with both MATLAB and \LaTeX\ can learn \Spike\ in a few hours.
\Spike\ is free, %: no logins, no passwords, no viruses or hidden malware:
MATLAB code of \Spike\ is open and is given in an appendix to this manual.

\bigskip
Example. The following problem is taken from D.\,C.\,Giancoli, {\it Physics for scientists and engineers.} 

\smallskip
\noindent
{\it A $9.0$ V battery whose internal resistance $r$ is $0.50\ \Omega$
is connected in the circuit shown in Figure below. How much current is drawn
from the battery?}

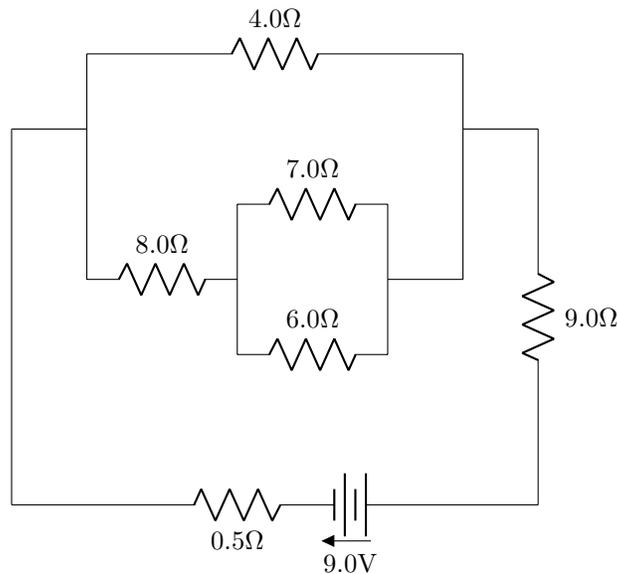
\begin{figure}[h!]
  \begin{center}
    \begin{circuitikz}
      \draw (0,0) to[short] (0,5) to[short] (1,5) to[short] (1,6)
      to[R=$4.0 \Omega$] (6,6) to[short] (6,5) to[short] (7,5)
      to[R=$9.0 \Omega$] (7,0) to[short] (5,0)
      to[battery=$9.0$V] (4,0)
      to[R=$0.5\Omega$] (2,0) to[short] (0,0);
      \draw (1,5) to[short] (1,3) to[R=$8.0 \Omega$] (3,3) to[short] (3,4) 
      to[R=$7.0 \Omega$] (5,4) to[short] (5,3) to[short] (6,3) to[short] (6,5);
      \draw (3,3) to[short] (3,2) to[R=$6.0 \Omega$] (5,2) to[short] (5,3);
    \end{circuitikz}
  \end{center}
\end{figure}

\medskip

To draw electric circuits we use the \verb!circuitikz! package by M.\,Redaelli.
The following spike-code generates randomised versions of this problem. 

{\small
\begin{verbatim}
<<problemR;
  R1,R2,R3,R4,R5!={3:15};
  V={7:20};
  r={0.3:0.1:0.7};
  R=R2+R3*R4/(R3+R4);
  R=R*R1/(R+R1);
  R=r+R+R5;
  I = V/R;
  answer('#2r',I);
@2;V, r, R1, R5, V, r, R2, R3, R4;
A $#1r$ V battery whose internal resistance $r$ is $#2r\ \Omega$
is connected in the circuit shown in Figure below. How much current (in A)
is drawn from the battery?
\begin{figure}[h!]
  \begin{center}
    \begin{circuitikz}
      \draw (0,0) to[short] (0,5) to[short] (1,5) to[short] (1,6)
      to[R=$#1r \Omega$] (6,6) to[short] (6,5) to[short] (7,5)
      to[R=$#1r \Omega$] (7,0) to[short] (5,0)
      to[battery=$#1r$V] (4,0)
      to[R=$#1r\Omega$] (2,0) to[short] (0,0);
      \draw (1,5) to[short] (1,3) to[R=$#1r \Omega$] (3,3) to[short] (3,4) 
      to[R=$#1r \Omega$] (5,4) to[short] (5,3) to[short] (6,3) to[short] (6,5);
      \draw (3,3) to[short] (3,2) to[R=$#1r \Omega$] (5,2) to[short] (5,3);
    \end{circuitikz}
  \end{center}
\end{figure}
>>
\end{verbatim}
}

\problemtype{R}
\QuesA{C}{2}
A $14.0$ V battery whose internal resistance $r$ is $0.70\ \Omega$
is connected in the circuit shown in Figure below. How much current (in A)
is drawn from the battery?
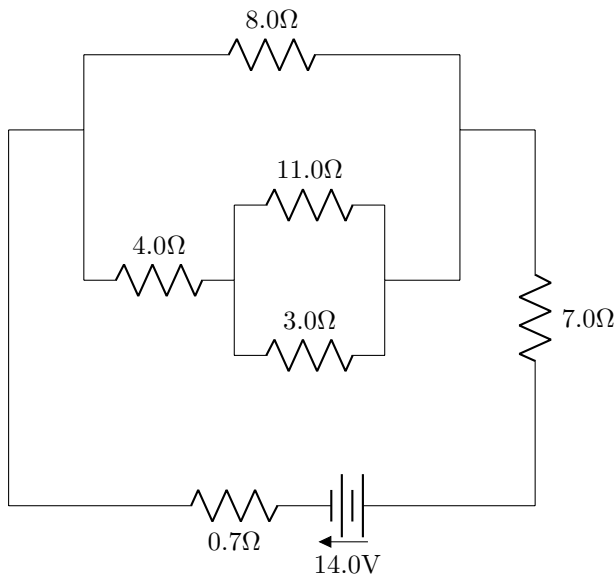
\begin{figure}[h!]
  \begin{center}
    \begin{circuitikz}
      \draw (0,0) to[short] (0,5) to[short] (1,5) to[short] (1,6)
      to[R=$8.0 \Omega$] (6,6) to[short] (6,5) to[short] (7,5)
      to[R=$7.0 \Omega$] (7,0) to[short] (5,0)
      to[battery=$14.0$V] (4,0)
      to[R=$0.7\Omega$] (2,0) to[short] (0,0);
      \draw (1,5) to[short] (1,3) to[R=$4.0 \Omega$] (3,3) to[short] (3,4)
      to[R=$11.0 \Omega$] (5,4) to[short] (5,3) to[short] (6,3) to[short] (6,5);
      \draw (3,3) to[short] (3,2) to[R=$3.0 \Omega$] (5,2) to[short] (5,3);
    \end{circuitikz}
  \end{center}
\end{figure}

\medskip\noindent
(A) \ $1.21$\quad
(B) \ $1.23$\quad
(C) \ $1.25$\quad
(D) \ $1.27$\quad
(E) \ $1.29$\quad

\vskip 1cm \section{How does \Spike\ work?}

%\subsection{Installing Spike?}
%\label{SS: How to obtain Spike}

\subsection{Installing Spike}
To install \Spike\ unzip the file 
\verb!spike.zip! to \verb!Matlab\! folder. 
The file \verb!spike.zip! contains Spike's Matlab-code, an example of top-file,
lots of examples of spike-files, etc. 

\subsection{To get started}
\Spike\ is written in MATLAB, and spike files are executed by MATLAB's 
processor. The aim of this program is to produce a LaTeX file which contains 
randomised assignments, one per each student.

An assignment has two attributes: assignment's topic and assignment number.
Hence, \Spike\ has two input arguments: topic code and assignment number. 
Assuming that the topic code is \verb!MATH3731! and the assignment being prepared is the third one,
%Assuming that all \Spike\ files have been properly installed,
the following Matlab command prepares the assignment:
\begin{verbatim}
>>spike('MATH3731',3)
\end{verbatim}
This command should be run from the folder \verb!\Spike\!, otherwise an error message is produced. 
The current folder \verb!\Spike\! should have 
a subfolder whose name coincides with the topic code, that is, \verb!MATH3731\!.
This folder will be referred to as the \emph{topic folder}.
The topic folder should have the following subfolders: 
\begin{verbatim}
  Final\
  Logs\
  Marks\
  Prelim\
  SPK\
  Submissions\
\end{verbatim}
\Spike\ saves the output LaTeX file with the assignment in the subfolder \verb!\Prelim\!. 
The output LaTeX file is named \verb!ANV1.tex!, so in this case it will be 
\verb!A3V1.tex!.

The file \verb!A3V1.tex! contains one assignment for each student. 
The topic folder should contain a file with a list of students' ID's enrolled in the topic.
The name of this file should be \verb!StudList.txt!. 
This file will be referred to as \emph{student list file}. It may look as follows:
\begin{verbatim}
2006280
2032971
2062720
2084704
-------
\end{verbatim}
These are student IDs of four students enrolled in the topic \verb!MATH3731!. 
As a result, \Spike\ will prepare four assignments. A file with list of 
students should have a line which starts with ``seven dashes'' \verb!-------!.
Everything which follows after this line is ignored by \Spike. This is 
convenient if there is a very large number of students enrolled in the topic, 
so in the testing regime one can temporarily put ``seven dashes'' after say the second line. 

The topic folder should also have a file which contains the \Spike-code 
for the assignment. The name of this file has strict format: \verb!AN.spk!.
So, in our case the topic folder must have a file \verb!A3.spk!.
This manual explains how to prepare these kind of \verb!spk!-files.

The current folder should also have a file with name \verb!topfile.tex!. This 
file will be called \emph{top-file}. \index{top-file}
The top-file contains assignment information such as due date, submission instructions, 
definitions of some \LaTeX\ commands, etc.

For example, the spike folder \verb!\Spike\! in my computer contains the following folders and files:

{\small
\begin{verbatim}
[+] MANUAL\
[+] MATH2702\
[+] MATH2722\
[+] MATH3712\
[-] MATH3731\
    [+] Final\
    [+] Logs\
    [+] Marks\
    [-] Prelim\
           A1V1.tex
           A2V1.tex
           A3V1.tex
           A4V1.tex
           A1V2.tex
           A2V2.tex
           A3V2.tex
           A4V2.tex
    [+} SPK       
    [+] Submissions\
        A1.spk
        A2.spk
        A3.spk
        A4.spk
        Ex.spk
        StudList.txt
[+] Spike_m_files\
[+] SpikeMark\
[+] STAT0000\
[+] Utils\
    examtopfile.tex
    spike.m
    examtopfile.tex
    topfile.tex
    what_is_new.txt
\end{verbatim}
}

\Spike\ overwrites an old assignment
\LaTeX-file if there is one.
Hence, as long as one is happy with the final version of the assignment 
and the assignment is ready to be posted on a university online learning site,
it is desirable if not necessary to copy the final version 
to a secure place where \Spike\ does not poke its nose.
The folder \verb!Final! is that secure place.

\smallskip

On successful completion of its work \Spike\ gives 
a message of the following kind:

{\small
\begin{verbatim}
>> spike('MATH3731',3)
spike, version 2017 D7, by Nurulla Azamov
Flinders University, Adelaide, Australia, 2014.

Preparing version 1 of the assignment:
Variant 1: 123456789012345678901234567890  done.
Variant 2: 123456789012345678901234567890  done.
...
Variant 21: 123456789012345678901234567890  done.
Version 1 is done.

Number of questions: 20
Number of problems: 43

Number of WARNINGS: 3.
The assignment file \MATH3731\Prelim\A3V1.tex has been copied to
..\..\Nur\Teaching Semester 2\MATH3731\Assignments
\end{verbatim}
}

Here \Spike\ prepared 21 variants of an assignment since apparently 21 students were enrolled in the topic
\verb!MATH3731!. Each assignment contains 30 questions.

In the case something goes wrong
\Spike\ produces an error message and allows to choose whether to continue
or to terminate the program.  
If something goes really wrong then \Spike\ will terminate preparation of the assignment and will 
try to explain what went wrong. For example:

{\small
\begin{verbatim}
spike, version D7, by Nurulla Azamov
Flinders University, Adelaide, Australia, 2014.

Preparing version 1 of the assignment:
Variant 1: 123456789012345678901


Spike: ERROR in spike-block starting at line 166 in spk-file.
Spike-block type: problemI
Line with error:
      !answer(#s,answ);!
Message:
      An answer command must have exactly two quotes.
Press Ctrl+Pause to terminate the program.
\end{verbatim}
}

%From the way the line of numbers \verb!12345678...! has been terminated after \verb!Variant 1:!
% one 
It can be seen that the cause of the error was in the 12th question.
In this particular case the error is that a line of code was missing quotes: it should be \verb!answer('#s',answ);!.

% \subsection{How \Spike\ works?}
% Spike does the following:
% \begin{enumerate}
%   \item It sets default values of certain parameters. These parameters can be changed in the top-file,

%   \item processes student ID's from the student list file,

%   \item creates for writing the out-file \verb!TopicCodeAN.tex!; in our case it is \verb!MATH3731A3.tex!,

%   \item opens the top-file and executes MATLAB commands which precede the line starting with \verb!\documentclass!,

%   \item copies the LaTeX code from the top-file into the out-file,

%   \item creates the assignment by processing spk-file.
% \end{enumerate}

\subsection{Disclaimer}

Copyright 2014 Nurulla Azamov.

Permission to use, copy and distribute this manual and software 
for any purpose is hereby granted free of charge, provided that the above 
copyright notice and this permission notice appear in all copies of this software and related documentation. 
This software is provided 'as-is' without any warranty of any kind. 

\section{Randomising variables}
To generate a random number one can use standard Matlab functions, such as 
\begin{verbatim}
  a = randi(10);
\end{verbatim}
Spike provides a different more flexible way of generating random numbers.

The command 
\begin{verbatim}
  p = {2,3,5,7,11,13,17,19};
\end{verbatim}
assigns to \verb!p! one of the prime integers $2,3,\ldots,19$ with equal probabilities. 
The command 
\begin{verbatim}
  a = {2:3,5:8,10:24,26:35};
\end{verbatim}
assigns to \verb!a! a random value from the range $1..35$ which is not a complete square,
and 
\begin{verbatim}
  b = {2:7,9:26,28:63,65:100};
\end{verbatim}
assigns to \verb!b! a random value from the range $1..100$ which is not a complete cube.
If one needs a larger range such as $1..10000,$ then 
the same goal in a spike-block can be achieved by 
\begin{verbatim}
  b = {2:10000};
  condition=(mod(b^(1/3),1)<0.001);
\end{verbatim}

The intervals may overlap, and the probability of a particular value is proportional  
to the number of times the value is listed. 
For instance, the command
\begin{verbatim}
  c = {0,0,1,1,1};
\end{verbatim}
assigns to \verb!c! the value $0$ with probability $0.4$ and the value $1$ 
with probability $0.6.$
Further, inside the curved brackets we can use Matlab formatted arrays, 
but in this case we should write \verb!:=! instead of \verb!=!, such as 
\begin{verbatim}
  d := {zeros(1,17),ones(1,23),2*ones(1,60)};
\end{verbatim}
This assigns to \verb!d! the value $0$ with probability $0.17,$
the value $1$ with probability $0.23,$ 
and the value $2$ with probability $0.60.$

One can use more than one pair of curved brackets in one command. For example,
\begin{verbatim}
  x = {1:5} + {3:6};
\end{verbatim}
This will assign to \verb!x! the sum of two random numbers 
from the ranges $1..5$ and $3..6$ respectively. 
The operator
\begin{verbatim}
  x = [{-3:5}, {3:6}, {-9:-2}];
\end{verbatim}
would return a random vector. 

The curved brackets can be nested. For example, one can use the operator
\begin{verbatim}
  t = {1{2:5}:3{3:5}};
\end{verbatim}
In such a case \Spike\ first processes inner pairs of curved brackets.

\subsection{Multiple assignment operators {\tt !=, !<, :=} and {\tt :<}}
In creating randomised questions it is often necessary to obtain two or more different random numbers
from the same range. There are four types of multiple assignment operators 
\verb!:=!, \verb!:<!, \verb~!=~ and \verb~!<~ which allow to do this.

For example, the command
\begin{verbatim}
  a,b!={1:8};
\end{verbatim}
assigns to \verb!a! and \verb!b! different values from the range $1..8.$ 
Here before the assignment operator \verb.!=. we can give a list of variables separated by commas. 
For each variable \verb!x! in the list
the assignment operator \verb!x={1:8}! will be executed in such a way that all variables 
in the list will get different random values.
It is also often desirable to have different values of variables which are sorted.
In this case one should use \verb.!<. instead of \verb.!=.. For instance, the command
\begin{verbatim}
  a,b,c,d!<{1:20};
\end{verbatim}
will assign different values to variables \verb!a,b,c,d! in such a way that \verb!a<b<c<d.!

The operators \verb.!=. and \verb.!<. we shall call multiple assignment operators.

The only restriction in the usage of multiple assignment operators 
is that there should be only one pair of curved brackets with some text between them and nothing else.
For example, the operators
\begin{verbatim}
  a,b,c,d!={1:7, 2:10, 13, 14};
  x,y,z!<{-0.1:0.7, 2:2:10};
\end{verbatim}
are ok, but the operators
\begin{verbatim}
  a,b,c,d!={1:7, 2:10, 1{3:4}};
  x,y,z!<{-1:7}/10;
\end{verbatim}
are not.

The array inside curved brackets does not need to be a number array.
For instance,
\begin{verbatim}
  a,b!={'A':'Z'};
  x,y,z!<{'a':'k', 'hello'};
\end{verbatim}
would work. In the second case here the three characters will be sorted in alphabetical order.

The operators \verb!:=! and \verb!:<! work as \verb:!=: and \verb:!<: with only one difference: 
the former pair of operators allows equal values for variables being assigned.

\section{Writing Spike-programs}
\subsection{Spike-blocks}
In this section we explain how to prepare a spike file.
A spike file consists of a sequence of spike blocks. A \emph{spike block} is a text which comes between \emph{spike-brackets} \verb!<<!
and \verb!>>!. Spike-brackets should be placed at the beginning of a line. If \verb!<<! or \verb!>>! is found somewhere inside a line,
then it is not considered as a spike-bracket.
The code in a spike-file is executed in a cycle $k$ times, where $k$ is the length 
of the student-list file. 

There are three types of spike-blocks: text-blocks, data-blocks and question-blocks.
In their turn question-blocks can be one of the following types: I, R, A, B, D, T, G, H and V.

A question-block has the format
{
\begin{verbatim}
<<problemX;
  ...
>>
\end{verbatim}
}
\noindent where \verb!X! is one of the letters I, R, A, B, D, T, G, H or V.
How to prepare a question-block will be explained in subsequent subsections.

\smallskip
% \Spike\ reads a spike file until it encounters the first spike block, 
% processes it and then proceeds reading the file until it finds the second 
% spike block, processes it and so on.
Any text outside spike-blocks is ignored by \Spike. 
Hence, one can use space between spike-blocks for comments.

Instead of a spike-block one can give a pointer to a file which contains a single spike-block,
in the following manner:
{
\begin{verbatim}
...
##[Path][filename]
...
\end{verbatim}
}
For example, 

{
\begin{verbatim}
<<text;
  \bigskip
  {\bf Elements of mathematicsl logic.}
>>

##MATH3731\SPK\Mathematical_logic\10.spk
##MATH3731\SPK\Mathematical_logic\20.spk
##MATH3731\SPK\Mathematical_logic\30.spk
##MATH3731\SPK\Mathematical_logic\40.spk
% ##MATH3731\SPK\Mathematical_logic\50.spk   not ready yet
##MATH3731\SPK\Mathematical_logic\60.spk
##MATH3731\SPK\Mathematical_logic\70.spk

<<text;
  \bigskip
  {\bf Sets.}
>>

##MATH3731\SPK\Sets\10.spk
##MATH3731\SPK\Sets\20.spk
##MATH3731\SPK\Sets\30.spk
##MATH3731\SPK\Sets\40.spk
\end{verbatim}
}

\subsection{Text-blocks}
If a spike-block is a text-block, then \Spike\ copies the content of the block into the output LaTeX file.

A text-block has the format
{
\begin{verbatim}
<<text;
  (Some text to be written
   into the output file)
>>
\end{verbatim}
}

For example, if we are about to prepare a spike-block which prepares a question about the Central Limit Theorem,
then we may write such a text-block:

{\small
\begin{verbatim}
<<text;
  \bigskip
  {\bf The Central Limit Theorem.}
>>
\end{verbatim}
}
One can write LaTeX commands in a text-block too, of course.

\subsection{Data-blocks}
A data-block has the format
{
\begin{verbatim}
<<data;
  (Matlab commands defining, say, some variables
   to be used in other spike-blocks)
>>
\end{verbatim}
}

In a data block we can define some variables.
For example, assume that a question asks to convert a number from base 10 to base $b.$
We can define the base $b$ in a data-block as follows:
{\small
\begin{verbatim}
<<data;
  b = {2:8};
>>
\end{verbatim}
}
The command \verb!b = {2:8};! assigns to $b$ a random value from the interval $2..8.$
Then the variable \verb!b! can be used in all subsequent spike-blocks, until it is redefined or removed.

\subsection{Questions of type I, R and A}

Questions of types I, R and A (I for ``integer'', R for ``real'',
A for ``approximate'') have the following structure:

{\small
\begin{verbatim}
<<problemI;
  (Command part)
  answer(...);
@Mark[:NRepeats[:NAltAns]];[parameters;]
  (Text of question)
>>
\end{verbatim}
}
\noindent 
where \verb!problemI! identifier should be replaced by 
\verb!problemR! or \verb!problemA! for the other two types. 

There is little difference between these three types of questions, so we will call them type IR questions.
Here the (Command part) is the command part of a type IR question,
\verb!Mark! is a one-digit number which is treated as number of points given for a correct answer to the question,
\verb!NRepeats! is the number of repeats of the question,
and \verb!NAltAns! is the number of multiplce choice answers. 
(Text of question) is the text part of the question which may contain some randomised variables,
and (parameters) are parameters of the text part of the question. 
The parameters \verb!NRepeats! and \verb!NAltAns! are not compulsory.

The command part consists of a sequence of MATLAB commands, but some of those 
commands can be special \Spike\ commands which are translated to MATLAB 
commands
by \Spike\ before feeding them to MATLAB's processor. The aim of these 
commands is
\begin{enumerate}
 \item to randomise certain elements of the multiple choice question,
 \item to find the correct answer to the question,
 \item to generate \verb!NAltAns-1! alternative incorrect answers.
\end{enumerate}

In the current version of \Spike\ all MATLAB commands should be written in one line.
For example, you can use double-nested \verb!for!-cycle but you should fit them into a single line.
This is a bit inconvenient but since one can use MATLAB functions in the command part this is not a big problem.

The text part of type IR question is the text of the multiple choice question which is typed in \LaTeX\ format and some elements of which are replaced
by values of variables given in the command part.
For example, a question asking to add three integer numbers may have the following text part:

\begin{center} \tt
     Find the sum of numbers 47, \verb!#r! and \verb!#r!.
\end{center}

This means that the first of these three numbers is not randomised while the other two are.
The commands \verb!#r! and \verb!#r! are replaced by values of two variables which
are given in the second \verb!@!-parameter,
for example, \verb!a,b;!. This means that the first number \verb!#r! will be replaced 
by the value of the variable \verb!a!,
and the second number will be replaced by the value of the variable~\verb!b!.

A type I question may look as follows:

{\small 
\begin{verbatim} 
<<problemI;
  a,b:={10:99};
  c=47+a+b;
  answer('$#r$',c);
@2;a,b;
Find the sum of numbers 47, #r and #r.
>>
\end{verbatim}
}

Here \verb!a,b:={10:99};! is a command which assigns to the variables \verb!a! and \verb!b! 
integer values chosen randomly from the interval 10..99.

The command \verb!c=47+a+b;! solves the question in the sense that it finds the correct answer.
Every type I or R question (but not A, more about this later) 
must also have ``the answer command''; in the example above it is
\begin{verbatim}
   answer('$#r$',c);
\end{verbatim}

\noindent \Spike\ replaces this command by the MATLAB command
\begin{verbatim}
   answer=xprintf('$#r$','c');
\end{verbatim}
before feeding it to MATLAB's processor, where \verb!xprintf! is a MATLAB 
function which will be discussed in depth later.
Thus, the answer command produces a string variable \verb!answer!
which is later used as one of the multiple choice answers.

Apart from this, a type IR question also produces
\verb!NAltAns-1! alternative incorrect answers. This is where the difference 
between type I and type R questions comes in.
To produce incorrect answers \Spike\ runs the command part of type I question 
\verb!NAltAns-1! more times. \Spike\ makes sure that
all the \verb!NAltAns! multiple choice answers are different. If the next answer
is not ``new'' then \Spike\ runs the command part again until a new answer is 
produced.
If \Spike\ fails to produce a new answer after a certain number of attempts, 
it stops and gives an error message.
Finally, \Spike\ randomly permutes the \verb!NAltAns! multiple choice answers, but it does not permute
the answers if the question is of type R, since in this case the multiple choice answers are naturally random.

\medskip

Example.

{\small 
\begin{verbatim}
<<text;
\bigskip
\noindent {\bf The Central Limit Theorem}
>>

<<problemI; 
  n={101:300};
  k=round(n/2+{-11:-2,2:11});
  p=0.5;
  q=1-p;
  z=((k+0.5)-n*p)/sqrt(n*p*q);
  r=normcdf(z,0,1);
  answer('#4r',r);
@2;n,k;
A fair coin is tossed #r times.
Find approximately the probability that 
the number of tails will not exceed #r.
>>
\end{verbatim}
}

\medskip
\noindent {\bf The Central Limit Theorem}

\problemtype{I}
\QuesA{B}{2}
A fair coin is tossed 287 times.
Find approximately the probability that
the number of tails will not exceed 154.

\medskip\noindent
(A) \ 0.2782
\quad 
(B) \ 0.9030
\quad 
(C) \ 0.1789
\quad 
(D) \ 0.2235
\quad 
(E) \ 0.9898
\quad 

\bigskip

In a type R question the answer is supposed to be a real number, such as $13.7834801\ldots.$
In this case \Spike\ rounds the answer according to instructions given in the command part
and then replaces the last digit of the answer by four other digits of the same parity:
if the last digit is odd (even) then \Spike\ replaces the last digit by odd (even) digits.

For example, if in the code above we replace \verb!problemI! by \verb!problemR! then the result will be different:

\QuesA{D}{2}
A fair coin is tossed 233 times.
Find the probability that the number of tails will not exceed 114.

\medskip\noindent
(A) \ $0.3960$\quad
(B) \ $0.3962$\quad
(C) \ $0.3964$\quad
(D) \ $0.3966$\quad
(E) \ $0.3968$\quad

\medskip
Type R questions have two advantages compared to type I: smart-guessing the correct answer becomes more difficult
if not impossible and the code runs faster.

\smallskip

Example. 

\smallskip

{\bf Question 1.} {\it Prevalence of a medical condition in a population is equal to 0.15.
What is the probability that exactly 3 people from a random sample
of 9 people chosen from this population have the condition?}

\smallskip \noindent We will randomise three elements of this question: prevalence, the size of random sample
and the number of people from the sample having the condition. 

{\small \begin{verbatim}
<<problemR;
  n={7:11};
  k={2:5};
  prvl={9:28}/100;
  p=nchoosek(n,k)*prvl^k*(1-prvl)^(n-k);
  answer('#5r',p);
@3;prvl,k,n;
Prevalence of a medical condition in a population is equal to #2r.
What is the probability that exactly #r people from a random sample
of #r people chosen from this population have the condition?
>>
\end{verbatim}
}

\problemtype{R}
\QuesA{D}{3}
Prevalence of a medical condition in a population is equal to 0.22.
What is the probability that exactly 3 people from a random sample
of 8 people chosen from this population have the condition?

\medskip\noindent
(A) \ $0.17210$ \quad
(B) \ $0.17212$ \quad
(C) \ $0.17214$ \quad
(D) \ $0.17216$ \quad
(E) \ $0.17218$ \quad

\medskip Everything will work in the same way if we replace the identifier \verb!problemR! by \verb!problemI!
but in this case the incorrect answers will be different.

\subsection{{\sf freeze} command in type I questions}
In type I questions Spike runs the command part \verb!NAltAns! times,
--- the first time to produce a question and the correct answer and the other \verb!NAltAns-1!
times to produce incorrect answers. Sometimes it is desirable to have a randomised 
variable, which would have the same value in all these \verb!NAltAns! runs. 
For this purpose there is \verb!freeze! command. All commands which appear before the \verb!freeze! command
are executed only once in the first run, in the other \verb!NAltAns-1! runs the same value is used. 

\medskip
Example. Here we produce a randomised system of two linear equations with coefficients 
from a field~$\mbZ_p.$ The value of $p$ needs to be frozen. 

{\small
\begin{verbatim}
<<problemI;
  p={7,11,13,17};
  freeze
  a,b,A,B!={2:p-1};
  condition = NotZero(mod(a*B-A*b,p));
  x,y!={2:p-1};
  c=mod(a*x+b*y,p);
  C=mod(A*x+B*y,p);
  rmdr=mod(x^2+y^2,p);
  answer('#r',rmdr);
@3:3;a,b,c,A,B,C,p;
The system of equations $#r x + #r y = #r $ and $#r x + #r y = #r $ 
has a unique solution $(x,y)$ in~$\mbZ_{#r}.$ Find this solution 
and then compute $x^2+y^2.$ 
>>
\end{verbatim}
}

\problemtype{I}
%Q2(a)
\QuesA{D}{3}

\smallskip
(a) 
The system of equations $3 x + 9 y = 12 $ and $13 x + 4 y = 4 $
has a unique solution $(x,y)$ in~$\mbZ_{17}.$ Find this solution
and then compute $x^2+y^2.$ 

\medskip\noindent
(A) \ 10\quad
(B) \ 1\quad
(C) \ 15\quad
(D) \ 8\quad
(E) \ 0\quad
%Raw volume 2.75e+07

\QuesB{B}(b) 
The system of equations $6 x + 2 y = 9 $ and $3 x + 7 y = 6 $
has a unique solution $(x,y)$ in~$\mbZ_{13}.$ Find this solution
and then compute $x^2+y^2.$ 

\medskip\noindent
(A) \ 2\quad
(B) \ 12\quad
(C) \ 11\quad
(D) \ 0\quad
(E) \ 6\quad

\QuesB{B}(c)\ 
The system of equations $10 x + 2 y = 3 $ and $3 x + 8 y = 10 $
has a unique solution $(x,y)$ in~$\mbZ_{11}.$ Find this solution
and then compute $x^2+y^2.$ 

\medskip\noindent
(A) \ 9\quad
(B) \ 4\quad
(C) \ 8\quad
(D) \ 5\quad
(E) \ 7\quad

\subsection{Questions of type A}
Sometimes a question requires to find an approximate solution of a problem.
In this case one may choose to use type $A$ question. 
The difference of type $A$ question from type $R$ one 
is that the command part in type $A$ question produces two variables
\verb!approxans! and \verb!exactans!, instead of the variable \verb!answer!.
To produce multiple choice answers Spike 
will choose \verb!approxans! with that number of digits which makes 
it different from \verb!exactans! by three digits. 

\medskip
Example.

{\small
\begin{verbatim}
<<problemA;
  c,d:={2:10}; 
  approxans=BisectionMethodF3(c,d,1,2,100,10^(-3));
  exactans=BisectionMethodF3(c,d,1,2,100,10^(-8));
@4:2;c,d;
Find a solution of the given equation in $[1,2]$ using 
the Bisection Method with tolerance $\verb!tol!=10^{-3}.$ 
As a stopping condition use $|f(x_n)| < \verb!tol!.$

\freeze
$x^5-#rx-#r=0$ 
>>
\end{verbatim}
}

\problemtype{A}
%Q17(a)
\QuesA{E}{4}\ \ Find a solution of the given equation in $[1,2]$ using
the Bisection Method with tolerance $\verb!tol!=10^{-3}.$
As a stopping condition use $|f(x_n)| < \verb!tol!.$

\medskip(a)\ $x^5-4x-4=0$

\medskip\noindent
(A) \ $1.5969841$\quad
(B) \ $1.5969843$\quad
(C) \ $1.5969845$\quad
(D) \ $1.5969847$\quad
(E) \ $1.5969849$\quad

%Raw volume 81

%Q17(b)
\QuesB{B}(b)\ $x^5-7x-2=0$

\medskip\noindent
(A) \ $1.6912840$\quad
(B) \ $1.6912842$\quad
(C) \ $1.6912844$\quad
(D) \ $1.6912846$\quad
(E) \ $1.6912848$\quad

\subsection{Volume of questions of type I, R and A}
With every question of type I, R or A we can associate a positive
integer which we shall call \emph{volume} of the question. By definition, the volume of a
question of type IR is the power of the set of all possible
different questions which \Spike\ can produce using the question's
code. There is also a second and different way to define the
volume: it is the power of the set of all possible different
answer strings which \Spike\ can produce using the question's code.
Often there is no difference between these two volumes but in
general they are different. To distinguish them, we shall call the
second volume the \emph{answer volume}. In any case the volume is
always $\geq$ the answer volume.

\medskip
Example.

{\small 
\begin{verbatim} 
<<problemI;
  a,b:={1:6};
  c=a+b;
  answer('$#r$',c);
@2;a,b;
Find the sum of numbers #r and #r.
>>
\end{verbatim}
}

\noindent The volume of this question is 36 and the answer volume is 11.

It is desirable to keep volumes of questions of type IR large
enough, say, at least 200. This is especially important for
questions of type I since small volumes slow down
processing of spike-code. More crucial reason for having a
large volume is to decrease a chance that two students
will get the same question.

Spike tries to estimate volume of a question and this estimate is shown after the question 
itself in the output LaTeX file. If the volume is small, Spike gives a warning in the
log-file. 

\medskip
Volume of a question is a measure of the size of the sample space of randomised variations of a question,
as such it does not evaluate the entropy of the corresponding probability distribution. 
Therefore it is possible that a type IR block with large volume but small entropy will often produce 
identical questions. 

\subsection{The variable {\tt condition} for questions of type I, R and A}
It is often necessary to ensure that random variables satisfy a certain contidition(s). 
For example, if a question asks to solve a system of two linear equations
with two unknowns then the coefficients $a,b,c,d$ of the system must satisfy the condition $ad-bc\neq 0.$
There is a special variable \verb!condition! in questions of type IR which facilitates this task.

The variable \verb!condition! may appear several times in one spike-block.
\Spike\ runs the command part in a cycle until all conditions hold. 

\medskip
Example. 

{\small
\begin{verbatim}
<<problemI;
  a1,a2!={2:15};
  b1,b2!={2:15};
  c1,c2!={2:15};
  condition=(a1*b1~=a2*b2);
  x=(a2*c2+a1*c1)/(a2*b2-a1*b1);
  answer('$#r$',x);
@1;a1,b1,c1,a2,b2,c2;
Solve the equation
$$
  #r(#r x + #r) = #r(#r x - #r).
$$
>>
\end{verbatim}
}

\QuesA{E}{1}
Solve the equation
$$
  7(13 x + 6) = 4(15 x - 7).
$$

\medskip\noindent
(A) \ $\frac{13}{3}$ \quad 
(B) \ $\frac{166}{27}$ \quad 
(C) \ $-\frac{203}{138}$ \quad 
(D) \ $45$ \quad 
(E) \ $-\frac{70}{31}$ \quad 

\subsection{The text part of questions of type I, R and A}
The text part of a question of type IR is what comes between the at-line and the closing spike-bracket \verb!>>!.
The text part is a plain text written in LaTeX format with one exception: the hash-tag symbol \verb!#!
has a special role. In the text part the hash-tag symbol is used 
as a formatting operator in a way similar to the usage of the percentage symbol in the Matlab function
\verb!fprintf!.

\medskip
Example.

{\small
\begin{verbatim}
<<problemI;
  a,b:={1:6};
  c=a+b;
  answer('$#r$',c);
@2;a,b;
Find the sum of numbers $#r$ and $#r.$
>>
\end{verbatim}
}

\subsection{Questions of type B}
If an answer to a question is a value of a categorical variable,
then it is preferable to use spike-block of type B.
The command part of type B block has to produce 
a discrete variable \verb!bool! (for ``boolean'') with values in \verb!0:N-1!, where \verb!N!
is the number of values of the categorical variable. 
The variable \verb!bool! indicates which of those values represents the correct answer:
if \verb!bool==0!, then the correct answer is (A), 
if \verb!bool==1!, then the correct answer is (B), etc. 

The values of the categorical variable should be listed in the \verb!answer! command,
separated by the tilde \verb!~! symbol. 
Type B block treats $N$ as the number of alternative multiple choice answers 
and therefore it ignores the third parameter of the at-line if there is one. 

\medskip 
Examples. 

{\small
\begin{verbatim}
<<problemB;
  [f,g]=GetIrreducibleDegree3Poly(p);
  bool={0,1};
  if bool==1, poly=f; else poly=g; end
  answer('No~Yes');
@2:2;p,poly;
Is the quotient ring $\mbZ_{#r}[x]/\langle #>p \rangle$ a field?
>>
\end{verbatim}
}

\QuesA{A}{2}
Is the quotient ring $\mbZ_{7}[x]/\langle 6x^{3}+5x^{2}+5x+6 \rangle$ a field?

\medskip\noindent
(A) \ No  \ \quad 
(B) \ Yes \ \quad  

\medskip
Here \verb!GetIrreducibleDegree3Poly(p)! is a Matlab function which generates a random pair of irreducible 
and reducible polynomials. 

\medskip
{\small
\begin{verbatim}
<<problemB;
  A=randi(11,2,2)-4;
  condition=NotZero(det(A));
  lmb=eig(A);
  a=real(lmb);
  condition=NotZero(a(1)*a(2));
  b=imag(lmb);
  nonreal=NotZero(b(1));
  arereal=~nonreal;
  if arereal && a(1)<0 && a(2)<0, bool=0; end
  if arereal && a(1)>0 && a(2)>0, bool=1; end
  if arereal && a(1)*a(2)<0, bool=2; end
  if nonreal && a(1)<0,  bool=3; end
  if nonreal && a(1)>0,  bool=4; end
  answer('stable\;node~unstable\;node~saddle~stable\;focus~unstable\;focus');
@2:2;A(1,:),A(2,:);
Find the type of zero solution of the following 
system of ODE's:
\begin{equation*}
  \begin{split}
     x' & = #"~x,y"l \\
     y' & = #"~x,y"l.
  \end{split}
\end{equation*}
>>
\end{verbatim}
}

\QuesA{B}{2}
Find the type of zero solution of the following
system of ODE's:
\begin{equation*}
  \begin{split}
     x' & = x+5y \\
     y' & = -x+7y.
  \end{split}
\end{equation*}

\medskip\noindent
(A) \ stable\;node    \ \quad 
(B) \ unstable\;node  \ \quad 
(C) \ saddle          \ \quad 
(D) \ stable\;focus   \ \quad 
(E) \ unstable\;focus \ \quad 

% One of the advantages of type B block compared to type I block is that type B block does not make Spike
% to run the command part several times to produce alternative answers. 

\subsection{Questions of type T}
A type T question (T for ``true'') has the following structure:
{\small
\begin{verbatim}
<<problemT;
% False statements
False statement 1 
False statement 2
...
False statement N
-------
% True statements
True statement 1
True statement 2
...
True statement M
@Mark[:NRepeats];k;
How many of the following assertions are true?
>>
\end{verbatim}
}

Here \Spike\ chooses randomly \verb!k! statements from the list of \verb!N+M! 
statements, where the first \verb!N! statements are false
and the next \verb!M! statements are true. True statements are given 
after the false ones, separated by the seven dashes \verb!-------!. 
The lines which start with \verb!%! are comments and are ignored.
Hence, the lines \verb!% False statements! and 
\verb!%True statements! are not necessary, but they are a good reminder.
Type T block does not produce multiple choice answers, since the number of true statements
is itself treated as an answer. 

\medskip
Example.

{\small \begin{verbatim}
<<problemT;
% False statements
Antelope
Zebra
Giraffe
Rabbit
Hippopotamus
Elephant
-------
% True statements
Tiger
Wolf
Lion
Cheetah
Bobcat
Cat
Coyote
@2:2;4;
How many of the following animals are carnivores?
>>
\end{verbatim}
}

\QuesA{2}{2} \ 

\smallskip
(a) 
How many of the following animals are carnivores?
\begin{enumerate}
\item Tiger
\item Elephant
\item Hippopotamus
\item Bobcat
\end{enumerate}

%Raw volume 715

\QuesB{2}(b) 
How many of the following animals are carnivores?
\begin{enumerate}
\item Giraffe
\item Lion
\item Coyote
\item Antelope
\end{enumerate}

\medskip
A statement in type T question can be written in more than one line,
but in this case the second and subsequent lines should start with at least three
empty spaces, as in the example shown below.

\medskip Example. 

{\small \begin{verbatim}
<<text;
\bigskip
\noindent {\bf Properties of $E(X)$ and $\Var(X)$}
>>
%%%%%%%%%%%%%%%%%%%%%%%%%%%%%%%%%%%%%%%%%%%%%%%%%
%  Properties of E(X) and Var(X)
%%%%%%%%%%%%%%%%%%%%%%%%%%%%%%%%%%%%%%%%%%%%%%%%%
<<problemT;
% False
$x_1+\ldots+x_n=1.$
$E(XY) = E(X)E(Y).$
$E(X^2) = [E(X)]^2.$
$E(\abs{X}) = \abs{E(X)}.$
$E(g(X)) = g(E(X)).$
For any real number $c$ \ $\Var(cX) = c\Var(X).$
If $c$ is a constant then $E(c) = 0.$
If $c$ is a constant then $\Var(c) = c.$
$E(X) \geq 0.$
-------------------------------------------------------------------
% True
$p_1+\ldots+p_n=1.$
$E(X+Y) = E(X)+E(Y).$
For any number $c$ \ $E(cX) = cE(X).$
If $c$ is a constant then $E(c) = c.$
$E(X) = p_1x_1+\ldots+p_nx_n.$
$E(X^2) = p_1x_1^2+\ldots+p_nx_n^2.$
$E(g(X)) = p_1g(x_1)+\ldots+p_n g(x_n).$
If $X$ and $Y$ are independent 
   then $E(XY) = E(X)E(Y).$
If $X$ and $Y$ are independent 
   then $\Var(X+Y) = \Var(X) + \Var(Y).$
For any real number $c$ \ $\Var(cX) = c^2\Var(X).$
$\Var(X) = E(X^2)- [(E(X)]^2.$
$\Var(X) \geq 0.$
$E(X^2) \geq 0.$
If $c$ is a constant then $\Var(c) = 0.$
If $\Var(X) = 0$ then $X$ is 
   a constant random variable.
% -------------------------------------------------------------------
@1;8;
Let $X$ be a random variable which takes values $x_1,\ldots,x_n$
and only these values with respective probabilities $p_1,\ldots,p_n,$
let $Y$ be another random variable and let $g(x)$ be a function.
How many of the following assertions are always correct?
>>
\end{verbatim}
}

\noindent {\bf Properties of $E(X)$ and $\Var(X)$}

\problemtype{T}
\QuesA{C}{1}
Let $X$ be a random variable which takes values $x_1,\ldots,x_n$ and only these values
with respective probabilities $p_1,\ldots,p_n,$ let $Y$ be another random variable and let $g(x)$ be a function.
How many of the following assertions are always correct?

\begin{enumerate}
\item If $c$ is a constant then $\Var(c) = 0.$
\item For any number $c$ \ $E(cX) = cE(X).$
\item $E(X^2) = [E(X)]^2.$
\item $\Var(X) = E(X^2)- [(E(X)]^2.$
\item $E(X) \geq 0.$
\item If $X$ and $Y$ are independent then $E(XY) = E(X)E(Y).$
\item $E(g(X)) = g(E(X)).$
\item $E(X^2) = p_1x_1^2+\ldots+p_nx_n^2.$
\end{enumerate}

\subsection{Randomising statements of type T question}

A type T question has a set of false and true statements 
to choose randomly from. Often certain elements of such statements 
can be varied without affecting their false/true value. 
For example, in a statement of the kind 
\begin{verbatim}
    The numbers 23 and 43 are congruent modulo 8.
\end{verbatim}
we can replace 23 by any other number of the kind $23+8k$ with integer $k$
without affecting value of the statement itself.
In such a case \Spike\ allows to randomise elements of statements. 
In order to do this in the case of the statement above we can replace it by 
\begin{verbatim}
    The numbers #r and 43 are congruent modulo 8.@23+8*{0:20}
\end{verbatim}
In this case the operator \verb!#r! will be replaced by one of the numbers $23, 23+8, \ldots, 23+8\cdot 20.$
We can randomise the second number too, for example, as follows:
\begin{verbatim}
    The numbers #r and #r are congruent modulo 8.@23+8*{0:20},43+8*{-9:9}
\end{verbatim}

% When \Spike\ finds out that a statement of a type T question contains the \verb!@!-character,
% it does the following: (1) it evaluates the randomising curved brackets as usual,
% it takes the statement itself into quotes and it replaces the \verb!@!-character by comma.
% After that \Spike\ feeds the resulting string to \verb!xprintf! function.

% For example, in the case of a string above, \Spike\ replaces the curved brackets 
% \verb!{0:20}! and \verb!{-9:9}! by numbers randomly chosen from the respective arrays, 
% for instance, $7$ and $-5.$ Then it feeds to Matlab's processor the command
% \begin{verbatim}
%     s=xprintf('The numbers #r and #r are congruent modulo 8.','23+8*7,43+8*-5');
% \end{verbatim}
% and uses the string \verb!s! as the statement. 

After the \verb!@!-character one can use variables which were defined in a data block.
For example, after the data block

{\small
\begin{verbatim}
  p={13,17,19,23,29,31,37};
\end{verbatim}
}
\noindent 
we can use in a type T question a randomised statement of this kind:

{\small
\begin{verbatim}
  The number $#r^{#r}$ is congruent to 1 modulo $#r.$@{2:p-1},p-1,p
\end{verbatim}
}

\bigskip

Example.

{\small
\begin{verbatim}
<<problemT;
% False
$u_{xx} #+cu_{xy} #+cu_{yy} 
   = u^{#r}_x + \cos(x)u_y$@{-6:-1,1:6},{-6:-1,1:6},{-6:-1,2:6}
$u^{#r}_{xx} #+cu_{xy} #+cu_{yy} 
   = #cu_x + \cos(x)u_y$@{2:6},{-6:-1,1:6},{-6:-1,1:6},{-8:8}
$u_{xx} #+cu^{#r}_{xy} - #r u_{yy} 
   = u_x + \cos(x)u_y$@{-6:-1,1:6},{2:6},{-6:-2,2:6}
$u_{xx} + #ru_{xy} + u_{yy} - #ru_x + #ru_y + \cos u 
   = 0$@{2:16},{2:16},{2:16} 
$u_{xx} + #ru_{xy} + u_{yy} - #ru_x + #ru_y + \sin u 
   = 0$@{2:16},{2:16},{2:16} 
$u_{xx} + #ru_{xy} + u_{yy} - #ru_x + #ru_y + e^u 
   = 0$@{2:16},{2:16},{2:16} 
$u u_{xxx} #+cu_{yyy} = u_x + e^{#cx } u_y$@{-8:8},{-8:-1,1:8}
$#r u^2_{xx} + #r u_{xy} + #r u_{yy} - #r u_x + #r u_y + #r u 
   = 0$@{2:16},{2:16},{2:16},{2:16},{2:16},{2:16}
------------------------------------------------------------
% True
$(#rx^2 + #ry)u_{xx} + \sin #rx\ u_{yy} 
   = \cos #r x \ u_x + \cos(x+y) u_y + e^{x+y}$@{2:16},{2:16},{2:16},{2:16}
$u_{xxx} - #r u_{yyy} = u_x + e^{#r x} u_y$@{2:16},{2:16}
$u_{xxxx} - #r u_{yyyy} = #ru_x + #ru_y$@{2:16},{2:16},{2:16}
$u_t = #r u_{xx}$@{2:21}
$u_{tt} = #r u_{xx}$@{2:21}
@1:2;5;
How many of the following partial differential
equations are linear (whether homogeneous or not)?
>>
\end{verbatim}
}

\QuesA{1}{1} \ 

\smallskip
(a) 
How many of the following partial differential
equations are linear (whether homogeneous or not)?
\begin{enumerate}
\item $u_{xx} + 14u_{xy} + u_{yy} - 11u_x + 2u_y + \cos u
   = 0$
\item $u^{6}_{xx} -5u_{xy} -6u_{yy}
   = -7u_x + \cos(x)u_y$
\item $u_{xx} + 2u_{xy} + u_{yy} - 15u_x + 2u_y + e^u
   = 0$
\item $u_{xx} -u_{xy} -u_{yy}
   = u^{-4}_x + \cos(x)u_y$
\item $u_{xxx} - 11 u_{yyy} = u_x + e^{16 x} u_y$
\end{enumerate}

%Raw volume 1287

\QuesB{3}(b) 
How many of the following partial differential
equations are linear (whether homogeneous or not)?
\begin{enumerate}
\item $u_t = 8 u_{xx}$
\item $u u_{xxx} -5u_{yyy} = u_x + e^{2x } u_y$
\item $u_{tt} = 14 u_{xx}$
\item $u_{xxx} - 11 u_{yyy} = u_x + e^{16 x} u_y$
\item $u_{xx} +4u^{5}_{xy} - -3 u_{yy}
   = u_x + \cos(x)u_y$
\end{enumerate}

\subsection{Toggling statements in type T questions}
If in the list of statements in a type T question we wish to give 
a negation of another statement then this can be done as follows:
after the false or true statement put \verb!&&! and then the negation.
In this case \Spike\ will choose with equal probability one of those two statements.

Example. 

{\small
\begin{verbatim}
<<problemT;
% False statements
Tom Cruise
Leonardo DiCaprio
Tom Hanks
Brad Pitt&&Jennifer Aniston
Johnny Depp&&Kate Winslet
Dwayne Johnson
Samuel Jackson
Ryan Gosling
-------
Helen Hunt&&Jack Nicholson
Julia Roberts&&Richard Gere
Scarlett Johansson&&Will Smith
Meryl Streep
Natalie Portman
Audrey Hepburn&&Morgan Freeman
@2:2; 5;
How many of these people are women?
>>
\end{verbatim}
}

\problemtype{T}
%Q2(a)
\QuesA{3}{2}\ \ (a)\ How many of these people are women?
\begin{enumerate}
\item Dwayne Johnson
\item Julia Roberts
\item Meryl Streep
\item Audrey Hepburn
\item Tom Hanks
\end{enumerate}

%Raw volume 6188

%Q2(b)
\QuesB{1}(b)\ How many of these people are women?
\begin{enumerate}
\item Johnny Depp
\item Tom Hanks
\item Tom Cruise
\item Natalie Portman
\item Leonardo DiCaprio
\end{enumerate}

\subsection{Questions of type G}
Questions of type G (G for ``generate'') are similar to questions of type T.

A type G question has the following structure 

{\small \begin{verbatim}
<<problemG;
  a_matlab_function_name
@Mark[:NRepeats[:NAltAns]];k;
How many of the following statements are true?
>>
\end{verbatim}
}
Here \verb!a_matlab_function_name! should be a MATLAB function which returns
two variables: a string variable which contains a statement in \LaTeX\ format
and a false/true variable, which indicates whether the statement is false or true.

In a question of type T we prepare a bank of false and true statements related to a certain topic.
But imagine, for example, that we want to test whether a student has learnt definition of congruence
modulo an integer.
We may ask him/her a question of this kind: ``Is it true that $\congruent{7}{38}{12}.$''
And we can prepare a bank of false and true statements of this kind for a type T question.
There is another way to do this: write a MATLAB function, say \verb!GetTwoCongruentIntegers!
which returns two variables: a string of the kind $\congruent{7}{38}{12}$ but with randomly chosen numbers
and a false/true value which shows whether the randomly generated statement is false or true. Once this 
is done we can create a type G question as follows:

{\small
\begin{verbatim}
<<problemG;
  GetTwoCongruentIntegers
@1;4;
How many of the following congruences are true?
>>
\end{verbatim}
}

\bigskip
\Spike\ will run the MATLAB function \verb!GetTwoCongruentIntegers! four 
times to produce four statements counting the number of true statements 
along the way.

This is what this spike-code produces:

\QuesA{D}{1}
How many of the following congruences are true?

\begin{enumerate}
\item $\congruent{108}{101}{7}$
\item $\congruent{88}{152}{8}$
\item $\congruent{31}{69}{10}$
\item $\congruent{63}{75}{3}$
\end{enumerate}

\medskip
Example. 

{\small \begin{verbatim}
function [s,b]=GetEvenInteger
% Returns a random integer s; b is true, if s is even
% and b is false if otherwise.

x=100+ 2 * randi(100);
b=(rand<0.5);
if ~b, x=x+1; end
s=xprintf('$#r$','x');
end
\end{verbatim}
}

{\small \begin{verbatim}
<<problemG;
 GetEvenInteger
@1;5;
How many of the following integers are even?
>>
\end{verbatim}
}

\QuesA{E}{1}
How many of the following integers are even?

\begin{enumerate}
\item $120$
\item $170$
\item $179$
\item $271$
\item $107$
\end{enumerate}

\subsection{Questions of type H}

Questions of type H are very similar to questions of type G,
but provide more flexibility.

Sometimes we may prepare two or more questions using the same data.
For example, given numbers $17, 32$ and $4$ the following statements are equivalent:
$$
  17 \in [32]_4, \ \ [17]_4 = [32]_4, \ \ \congruent{17}{32}{4}.
$$
In such a case type $H$ spike-block allows to generate three different questions using the same data.
The advantage of this is that we need to write only one Matlab function producing randomised data,
such as $17,32,4,$ in order to generate two or more questions.

In type $H$ question the command part has to produce two variables: \verb!out! and \verb!tf! (for true/false).
The first variable \verb!out! is a string variable which contains in LaTeX format the statement of the question
and the second variable \verb!tf! is a boolean variable which indicates whether this statement is true or not.

The structure of a type H question is this:

{\small \begin{verbatim}
<<problemH;
  (Command part)
@Mark[:NRepeats];k;
How many of the following statements are true?
>>
\end{verbatim}
}

The command part has to produce two variables: \verb!out! and \verb!tf!.
\Spike\ runs the command part \verb!k! times to generate \verb!k! statements 
counting along the way the number of true statements produced.

Let us consider an example.
Assume that we have written a MATLAB function \verb!GetCongruentInts! which 
returns four variables: three random integer variables \verb!a!, \verb!b! and 
\verb!n!
and one boolean variable \verb!tf!, which has value ``true'' if and only if 
\verb!n! divides \verb!a-b!.
Then the aim of question 18 can be achieved by the following code:

{\small
\begin{verbatim}
<<problemH;
  [a,b,n,tf]=GetCongruentInts;
  out=xprintf('$\congruent{#r}{#r}{#r}$','a,b,n');
@1;4;
How many of the following statements are true?
>>
\end{verbatim}
}

The LaTeX command \verb!\congruent! can be defined as, for example, \\
\noindent
\verb!\newcommand{\congruent}[3]{#1 \equiv #2 \ \ (\,\mathrm{mod} \ #3)}!. \\

\noindent
This definition can be inserted into the top-file.

Now using the same function \verb!GetCongruentInts! we can generate other questions. For example,

{\small
\begin{verbatim}
<<problemH;
  [a,b,n,tf]=GetCongruentInts;
  out=xprintf('$[#r]_{#r}=[#r]_{#r}$','a,n,b,n');
@1;4;
How many of the following statements are true?
>>

<<problemH;
  [a,b,n,tf]=GetCongruentInts;
  out=xprintf('$#r \in [#r]_{#r}$','a,b,n');
@1;4;
How many of the following statements are true?
>>

<<problemH;
  [a,b,n,tf]=GetCongruentInts;
  out=xprintf('$[#r]_{#r} \cap[#r]_{#r} = \\emptyset$','a,n,b,n');
@1;4;
How many of the following statements are true?
>>
\end{verbatim}
}

\problemtype{H}
\QuesA{D}{1}
How many of the following statements are true?
\begin{enumerate}
\item $[106]_{4}=[138]_{4}$
\item $[33]_{13}=[181]_{13}$
\item $[56]_{3}=[53]_{3}$
\item $[119]_{10}=[138]_{10}$
\end{enumerate}

\problemtype{H}
\QuesA{C}{1}
How many of the following statements are true?
\begin{enumerate}
\item $44 \in [132]_{7}$
\item $112 \in [180]_{5}$
\item $113 \in [146]_{3}$
\item $35 \in [178]_{13}$
\end{enumerate}

\problemtype{H}
\QuesA{C}{1}
How many of the following statements are true?
\begin{enumerate}
\item $[78]_{4} \cap[114]_{4} = \emptyset$
\item $[61]_{3} \cap[98]_{3} = \emptyset$
\item $[115]_{3} \cap[124]_{3} = \emptyset$
\item $[115]_{9} \cap[178]_{9} = \emptyset$
\end{enumerate}

\subsection{Questions of type D}

Example of a type D question:

{\small
\begin{verbatim}
<<problemD;
  Spike_m_files/ProblemD_data/animals.txt
  **1
@2;5;
How many of the following are cats?
>>
\end{verbatim}
}

Here \verb!Spike_m_files/ProblemD_data/animals.txt!
is a file with so-called problem D database. It has the following content:

{\small
\begin{verbatim}
% Carnivore: C carnivore, H herbivore
% Type: M mammal, F fish, R reptile, B bird.
% Cats: 1 a cat, 0 not a cat.
% Animal       Carnivore     Type       Cats
%
%
Antelope   &   H         &   M       &   0
Bobcat     &   C         &   M       &   1
Cheetah    &   C         &   M       &   1
Crocodile  &   C         &   R       &   0
Emu        &   H         &   B       &   0
Jaguar     &   C         &   M       &   1
Kangaroo   &   H         &   M       &   0
Koala      &   H         &   M       &   0 
Leopard    &   C         &   M       &   1
Lion       &   C         &   M       &   1
Panthera   &   C         &   M       &   1
Rabbit     &   H         &   M       &   0
Shark      &   C         &   F       &   0
Tiger      &   C         &   M       &   1
Wolf       &   C         &   M       &   0 
Zebra      &   H         &   M       &   0
\end{verbatim}
}
\noindent 
Lines which start with \verb!%! are comments. 
Spike picks up five (from the second parameter in the at-line \verb!@2;5;!)
entries of this database. Which of them are considered to be true is determined by the second line
\verb!**1!. The line \verb!**1! means that a chosen entry is correct if the third parameter is 1,
while the first two can be anything.

One can also tell Spike which statements to choose. For example, 

{\small
\begin{verbatim}
<<problemD;
  Spike_m_files/ProblemD_data/animals.txt
  **1
  *M*
@2;5;
How many of the following mammals are cats?
>>
\end{verbatim}
}
\noindent 
The third line \verb!*M*! here tells Spike to choose only entries which satisfy the 
specifications \verb!*M*!, that is, entries whose second parameter is \verb!M!, mammal.

\problemtype{D}
\QuesA{C}{2}
How many of the following mammals are cats?
\begin{enumerate}
\item Antelope
\item Rabbit
\item Cheetah
\item Jaguar
\item Panthera
\end{enumerate}

\bigskip

Example. 
Assume that we have prepared the following file \verb!rings.txt!:
{\small
\begin{verbatim}
%                & Ring    &  Commutative  & Has identity  & Has no zero divisors  & Has mult. inv.
$\mbZ$           & 1       &  1            & 1             & 1                     & 0 
$\mbQ$           & 1       &  1            & 1             & 1                     & 1
$\mbR$           & 1       &  1            & 1             & 1                     & 1
$\mbC$           & 1       &  1            & 1             & 1                     & 1
$C(\mbR)$        & 1       &  1            & 1             & 0                     & 0
$C_c(\mbR)$      & 1       &  1            & 0             & 0                     & 0
$C^1(\mbR)$      & 1       &  1            & 1             & 0                     & 0
$C^2(\mbR)$      & 1       &  1            & 1             & 0                     & 0
$C^\infty(\mbR)$ & 1       &  1            & 1             & 0                     & 0
$M_n(\mbR)$      & 1       &  0            & 1             & 0                     & 0
$\mbZ_2$         & 1       &  1            & 1             & 1                     & 1
$\mbZ_3$         & 1       &  1            & 1             & 1                     & 1
$\mbZ_5$         & 1       &  1            & 1             & 1                     & 1
$\mbZ_7$         & 1       &  1            & 1             & 1                     & 1
$\mbZ_{11}$      & 1       &  1            & 1             & 1                     & 1
$\mbZ_4$         & 1       &  1            & 1             & 0                     & 0
$\mbZ_6$         & 1       &  1            & 1             & 0                     & 0
$\mbZ_8$         & 1       &  1            & 1             & 0                     & 0
$\mbZ_9$         & 1       &  1            & 1             & 0                     & 0
$\mbZ_{10}$      & 1       &  1            & 1             & 0                     & 0
$\mbZ_{12}$      & 1       &  1            & 1             & 0                     & 0
-------
\end{verbatim}
}

{\small
\begin{verbatim}
<<problemD;
  Spike_m_files/ProblemD_data/rings.txt
  11111
@2;5;
How many of the following rings are fields?
>>
\end{verbatim}
}

\problemtype{D}
\QuesA{D}{2}
How many of the following rings are fields?
\begin{enumerate}
\item $\mbZ_8$
\item $\mbZ_5$
\item $C^1(\mbR)$
\item $\mbC$
\item $\mbZ_3$
\end{enumerate}

\bigskip 

{\small
\begin{verbatim}
<<problemD;
  Spike_m_files/ProblemD_data/rings.txt
  11111
  111**
@2;5;
How many of the following unital commutative rings are fields?
>>
\end{verbatim}
}

\problemtype{D}
\QuesA{E}{2}
How many of the following unital commutative rings are fields?
\begin{enumerate}
\item $C^2(\mbR)$
\item $\mbZ_7$
\item $\mbR$
\item $\mbC$
\item $\mbZ_{12}$
\end{enumerate}

\subsection{The $\mathrm{\backslash}${\tt freeze} command}
In the text part of all question-blocks one can use the command \verb!\freeze!.
This command is used if the question is repeated at least two times.
The part of the text part which appears before the \verb!\freeze! command
is typed only once, and the part which follows after \verb!\freeze! is typed 
in every repeat of the question. 

Examples. 

{\small
\begin{verbatim}
<<problemB;
  n={8:20};
  a = randperm(n);
  parity=permparity(a);
  bool=round((1-parity)/2);
  answer('$+1$~$-1$');
@2:3;a;
Find parity of the permutation.

\freeze
$$
  #0q.
$$
>>
\end{verbatim}
}

\problemtype{B}
%Q4(a)
\QuesA{B}{2} \ Find parity of the permutation.

\smallskip
(a) 
$$
  \left(
  \begin{array}{ccccccccccccccccc}
    1&2&3&4&5&6&7&8&9&10&11&12&13&14&15&16&17\\
    12&10&8&3&1&14&5&13&4&11&2&15&7&9&6&16&17
  \end{array}
\right).
$$

\medskip\noindent
(A) \ $+1$ \ \quad 
(B) \ $-1$ \ \quad 

%Raw volume 1.30e+07

\QuesB{A}(b) 
$$
  \left(
  \begin{array}{cccccccccc}
    1&2&3&4&5&6&7&8&9&10\\
    1&2&8&4&7&10&6&3&9&5
  \end{array}
\right).
$$

\medskip\noindent
(A) \ $+1$ \ \quad 
(B) \ $-1$ \ \quad 

\QuesB{B}(c)\ 
$$
  \left(
  \begin{array}{cccccccccc}
    1&2&3&4&5&6&7&8&9&10\\
    10&2&3&8&4&7&6&5&1&9
  \end{array}
\right).
$$

\medskip\noindent
(A) \ $+1$ \ \quad 
(B) \ $-1$ \ \quad

\bigskip
{\small
\begin{verbatim}
<<problemT;
% False
$\phi \colon \mbR \to \mbR,$ $\phi(x) = e^x.$
$\phi \colon GL(#r,\mbR) \to \mbR,$ \ $\phi(A) = \det(A).$@{2:6}
$\phi \colon M_{#r}(\mbR) \to \mbR^*,$ \ $\phi(A) = \det(A).$@{2:6}
$\phi \colon \mbR \to \mbR,$ $\phi(x) = 1.$
$\phi \colon M_n(\mbR) \to \mbR,$ \ $\phi(A) = \det(A).$
-------
% True
$\phi \colon GL(n,\mbR) \to \mbR^*,$ \ $\phi(A) = \det(A).$
$\phi \colon \mbR \to \mbR_+,$ $\phi(x) = e^x.$
$\phi \colon \mbR_+ \to \mbR,$ $\phi(x) = \ln(x).$
$\phi \colon \mbR \to \mbT,$ $\phi(x) = e^{i#rx}$@{2:15}
$\phi \colon \mbZ \to \mbT,$ $\phi(m) = e^{i#rm}$@{2:15}
$\phi \colon \mbR \to \mbR_+,$ $\phi(x) = 1.$
$\phi \colon \mbR \to \mbR,$ $\phi(x) = 0.$
@4:2;4;
In the following questions $\mbZ,\ \mbR$ are groups with addition 
as group operation, $\mbR_+:=\set{x \in \mbR \colon x>0},$ 
$\mbR^*:=\mbR\setminus\set{0}$ and $\mbT:=\set{z \in \mbC \colon 
\abs{z}=1}$ are groups with multiplication as group operation. 
$M_n(\mbR)$ is the set of all real matrices of size $n\times n$ 
with matrix addition as a group operation, $GL(n,\mbR)$ is the set 
of all invertible real matrices of size $n\times n$ with matrix 
multiplication as a group operation. 

\freeze
How many of the following 
maps are group homomorphisms?
>>
\end{verbatim}
}

\problemtype{T}
%Q3(a)
\QuesA{4}{4} \ In the following questions $\mbZ,\ \mbR$ are groups with addition
as group operation, $\mbR_+:=\set{x \in \mbR \colon x>0},$
$\mbR^*:=\mbR\setminus\set{0}$ and $\mbT:=\set{z \in \mbC \colon
\abs{z}=1}$ are groups with multiplication as group operation.
$M_n(\mbR)$ is the set of all real matrices of size $n\times n$
with matrix addition as a group operation, $GL(n,\mbR)$ is the set
of all invertible real matrices of size $n\times n$ with matrix
multiplication as a group operation.

\smallskip
(a) 
How many of the following
maps are group homomorphisms?
\begin{enumerate}
\item $\phi \colon \mbZ \to \mbT,$ $\phi(m) = e^{i5m}$
\item $\phi \colon \mbR_+ \to \mbR,$ $\phi(x) = \ln(x).$
\item $\phi \colon \mbR \to \mbR_+,$ $\phi(x) = 1.$
\item $\phi \colon \mbR \to \mbR,$ $\phi(x) = 0.$
\end{enumerate}

%Raw volume 4.95e+02

\QuesB{2}(b) 
How many of the following
maps are group homomorphisms?
\begin{enumerate}
\item $\phi \colon M_n(\mbR) \to \mbR,$ \ $\phi(A) = \det(A).$
\item $\phi \colon \mbR \to \mbR,$ $\phi(x) = e^x.$
\item $\phi \colon \mbR_+ \to \mbR,$ $\phi(x) = \ln(x).$
\item $\phi \colon GL(n,\mbR) \to \mbR^*,$ \ $\phi(A) = \det(A).$
\end{enumerate}

\subsection{Questions of type V}
A type V question is not random.
This type of question is introduced to allow non-randomised questions
in an assignment. The structure of a type V question is as follows:

{\small
\begin{verbatim}
<<problemV;
@Mark;X;
Write here whatever you want
in LaTeX format.
>>
\end{verbatim}
}

Here \verb!X! is the correct answer to the question. 

\medskip
Example.

{\small
\begin{verbatim}
<<problemV;
@2;D;
Which of the following chemical elements is a metal?
\medskip

 (A) \ $\mathrm O$ \quad  (B) \ $\mathrm C$ \quad  (C) \ $\mathrm H$
 \quad  (D) \ $\mathrm K$ \quad  (E) \ $\mathrm N$
>>
\end{verbatim}
}

\QuesA{D}{2}
Which of the following chemical elements is a metal?
\medskip

 (A) \ $\mathrm O$ \quad  (B) \ $\mathrm C$ \quad  (C) \ $\mathrm H$
 \quad  (D) \ $\mathrm K$ \quad  (E) \ $\mathrm N$

\subsection{Questions of type Z}
A question of type Z is not a question.
If a spike-block \verb!<<...>>! starts with \verb!<<problemZ;!
then \Spike\ ignores the whole spike-block. This feature is convenient
for toggling questions in and out from the assignment. 

\section{Top-file}

\subsection{Top-file}
\Spike\ prepares an output \LaTeX-file which contains one assignment for each student enrolled in a topic.
In the beginning of this output file it is desirable to give some information relevant to the assignment such as:
(1) assignment number, (2) due date, (3) the percentage of the total assessment which this assignment constitutes,
(4) instructions on how to submit answers to the assignment, (5) what to do in certain exceptional cases, etc.

All this information can be given in the top-file.
It can also contain \LaTeX\ preamble commands and macro-commands.
An example of a top-file is given in Section \ref{A: example of top-file}.

The command \verb!%EOF! in the top-file means end of file: \Spike\ ignores everything which follows after this line.
This command is not compulsory.

Top-file consists of two parts: the MATLAB part and the \LaTeX-part.
The line which starts with \verb!\documentclass! separates the MATLAB and the \LaTeX-parts of the top-file.
Spike executes commands in the MATLAB part and copies the \LaTeX-part into the beginning
of the output \LaTeX-file.

Some of the \LaTeX-part commands may contain references to the topic code and the assignment number, which are given as arguments to \Spike.
In \Spike\ there are two \LaTeX-commands \verb!\topiccode! and \verb!\NAss! which define the topic code and the assignment number respectively.
\Spike\ automatically defines these commands when it encounters the line \verb!\begin{document}! in the top-file.
That is, in the top-file one can use \LaTeX\ commands \verb!\topiccode! and \verb!\NAss! after \verb!\begin{document}! without defining them.

\subsection{\Spike\ variables}
Spike has several variables:
\verb!duedate, rseed, NAltAns, NRepeat.!
\verb!rseed! is a seed for MATLAB's generator of pseudo-random numbers, \verb!NAltAns! is the number of alternative answers,
\verb!NRepeat! is the number of times each assignment question is repeated.

By default, \verb!NAltAns!=5, and \verb!NRepeat!=1.
These default values can be changed in the MATLAB part of the top-file or in a data-block, as in 
\begin{verbatim}
  NAltAns = 4;
\end{verbatim}
or
\begin{verbatim}
<<data;
  NAltAns = 4;
>>
\end{verbatim}

\subsection{Destination folder {\tt DestFolder}}

Spike saves the assignment file in the \verb!\Prelim! folder.
One can also instruct Spike to save the assignment file
in an another folder, which we call \emph{destination folder}, if such a folder exists. 
The destination folder is defined in the top-file by assigning value to the \verb!DestFolder!
variable, for example,
\begin{verbatim}
DestFolder='..\..\Nur\Teaching Semester 2\MATH3731\Assignments';
\end{verbatim}
Spike starts searching for this folder from the \verb!Spike! folder. 

In the name of \verb!DestFolder! one can use the parameters \verb![Sem]!, \verb![Ver]!
and \verb![Topic]! which will be replaced by the current semester, 
assignment version number and the topic code, for example, as in 
\begin{verbatim}
DestFolder='..\..\Nur\Teaching Semester [Sem]\[Topic]\Assignments';
\end{verbatim}

\subsection{Variable {\tt duedate}}
In every run \Spike\ prepares a LaTeX command \verb!\DueDate!, for example,
\begin{verbatim}
\newcommand{\DueDate}{Wednesday, 14th of September, 11pm}
\end{verbatim}
Here the string \verb!'Wednesday, 14th of September, 11pm'! is taken from the value of the variable 
\verb!duedate!, which needs to be set up in the Matlab part of the top-file;
for example, 
\begin{verbatim}
duedate='Wednesday, 14th of September, 11pm';
\end{verbatim}
After this the LaTeX command \verb!\DueDate! can be used in the LaTeX part of the top-file. 

\smallskip Often it is convenient to set up due dates of particular assignments for particular topics
beforehand. This can be done as follows: 
\begin{verbatim}
duedate:MATH3712A2V1='Wednesday, 14th of September, 11pm';
\end{verbatim}
This means that version 1 of assignment 2 for topic \verb!MATH3712! is due on the specified day and time. 
This due date will be applied only to such an assignment, that is, only if the Matlab command 
which generates the assignment is \verb!spike('MATH3712',2)! or \verb!spike('MATH3712',2,1)!;
otherwise this command will be ignored by \Spike. There can be more than one \verb!duedate! commands. 
For example, the commands
{\small
\begin{verbatim}
% Due dates of Partial Differential Equations assignments. 
duedate:MATH3712A1V1='Wednesday, 24th of August, 11pm';
duedate:MATH3712A2V1='Wednesday, 14th of September, 11pm';
duedate:MATH3712A3V1='Wednesday, 19th of October, 11pm';
duedate:MATH3712A4V1='Friday, 4th of November, 11pm';

% Due dates of Algebra assignments. 
duedate:MATH3731A1V1='Wednesday, 24th of August, 11pm';
duedate:MATH3731A2V1='Wednesday, 14th of September, 11pm';
duedate:MATH3731A3V1='Wednesday, 19th of October, 11pm';
duedate:MATH3731A4V1='Friday, 4th of November, 11pm';
\end{verbatim}
}
set up the due dates of the eight assignments. 

\smallskip If there are more than one eligible \verb!duedate! commands, then \Spike\ 
will choose the last eligible \verb!duedate! command, ignoring all previous \verb!duedate! commands. 

\subsection{Variable {\tt PercentMarks}}
Everything that has been said in the previous subsection about the \verb!duedate! variable also 
applies to \verb!PercentMarks! variable. This variable is supposed to show the relative weight
of an assignment in the topic assessment. 
For example, if the Matlab part of the top-file has a command 
\begin{verbatim}
PercentMarks='10';
\end{verbatim}
then \Spike\ will create a LaTeX variable \verb!\PercentMarks! using the command 
\begin{verbatim}
\newcommand{\PercentMarks}{10}.
\end{verbatim}
This command can further be used in the LaTeX part of the top-file. 

\smallskip One can also set up weights of individual assignments. 
For example, if a topic \verb!MATH2702! has five assignments with maximum marks 5, 10, 5, 10 and 10 
respectively then one can use the following commands:

{\small
\begin{verbatim}
PercentMarks:MATH2702A1V1=5;
PercentMarks:MATH2702A2V1=10;
PercentMarks:MATH2702A3V1=5;
PercentMarks:MATH2702A4V1=10;
PercentMarks:MATH2702A5V1=10;
\end{verbatim}
}

\subsection{Variables {\tt GlblNRepeat, GlblWeight} and {\tt GlblNAltAns}}
There are variables called \verb!GlblNRepeat!, \verb!GlblWeight! and \verb!GlblNAltAns!.
By default, these variables have the value of the empty string, in which case they are ignored by \Spike.
If the value of any of these variables is not an empty string then this value overrides the value of the 
corresponding ``local'' variable. For example,
an at-line of this kind
{\small
\begin{verbatim}
@4:2:4; ...
\end{verbatim}
}
\noindent 
sets the weight of the question to be $4,$ the number of repeats to be $2$ and the number of alternative answers 
to be $4.$ However, these values will be overridden by the values of the corresponding variables 
\verb!GlblNRepeat!, \verb!GlblWeight! and \verb!GlblNAltAns!, provided they are values are not empty strings. 
The variables \verb!GlblNRepeat! and \verb!GlblNAltAns! also override values of variables 
\verb!NRepeat! and \verb!NAltAns!.

This feature is convenient for generating exam papers. In an exam paper 
the number of repeats of every question should probably be set to $1$ (right?), 
the weights of exam questions may be chosen to be the same independently of their difficulty, 
and the number of alternative answers can be chosen to be say four instead of normal five. 
It would be absolutely unwise to change manually values of these variables in all questions for this purpose.
The variables \verb!GlblNRepeat!, \verb!GlblWeight! and \verb!GlblNAltAns! provide a convenient way 
of doing this.  

For example, in my exam top-file \verb!examtopfile.tex! I have the following lines:
{\small
\begin{verbatim}
GlblWeight=3;
GlblNRepeat=1;
GlblNAltAns=5;
\end{verbatim}
}

The following is a fragment of a resulting exam paper: 

\newcommand{\QuesAZ}[1]{\bigskip \addtocounter{nques}{1} %
   \noindent \mbox{{\bf\arabic{nques}.}\,[$#1$m]}}

\problemtype{I}
%MATH2722\SPK\Polynomial_interpolation\osculating_polynomial.spk
%Q4(a)
\QuesAZ{3}\ \ Let $P_2(x)=ax^2+bx+c$ be the osculating polynomial
for the data $$f(-1)=-7, \ \ f(2)=-4, \ \ f\dash(-1)=7.$$
Find the coefficients of $P_2(x)$ and thereby calculate $a-2b+3c.$

\medskip\noindent
(A) \ $28$\quad
(B) \ $-14$\quad
(C) \ $-15$\quad
(D) \ $1$\quad
(E) \ $13$\quad
%Raw volume 594
%Corr ans: B

\problemtype{I}
%MATH2722\SPK\Numerical_differentiation\3point_leftend.spk
%Q5(a)
\QuesAZ{3}\ \ Find the relative error of approximation of $f\dash(0.6)$
  by the Left End 3-point formula with mesh size $h=0.015,$
  where $f(x) = \arctan(7x).$

\medskip\noindent
(A) \ $2.1536\cdot 10^{-4}$\quad
(B) \ $2.1566\cdot 10^{-3}$\quad
(C) \ $5.3853\cdot 10^{-4}$\quad
(D) \ $3.2704\cdot 10^{-4}$\quad
(E) \ $1.0271\cdot 10^{-3}$\quad
%Raw volume 1.27e+42
%Corr ans: E

\problemtype{R}
%Q6
\QuesAZ{3}\ \ Find the relative error of approximation of the integral
  $$\int_0^{0.4} \frac{1}{1+2x^2}\,dx$$
  by Trapezoidal Rule.

\medskip\noindent
(A) \ $8.6385\cdot 10^{-2}$\quad
(B) \ $3.2453\cdot 10^{-2}$\quad
(C) \ $3.4359\cdot 10^{-2}$\quad
(D) \ $1.6470\cdot 10^{-2}$\quad
(E) \ $1.6855\cdot 10^{-2}$\quad
%Corr ans: C

\section{Function {\tt xprintf} and hash-tag operators}
\label{S: hashtag op-rs}

In this section we describe the function {\tt xprintf},
which converts data from Matlab format into LaTeX format.

MATLAB has an in-built function \verb!sprintf!. 
For example,

{\small
\begin{verbatim}
>> a=4; b=5; 
>> sprintf('As is known %d times %d equals %d.',a,b,a*b)
ans =
As is known 4 times 5 equals 20.
\end{verbatim}
}

Spike has an analogue of this function called \verb!xprintf! which comes in \Spike's package. 
The difference between \verb!sprintf! and \verb!xprintf! is that the latter represents 
the variables in LaTeX format, and unlike \verb!sprintf!, in \verb!xprintf! the variables 
should be taken in quotes. 

For example, the same result as above is achieved by the command

{\small
\begin{verbatim}
>> xprintf('As is known #r times #r equals #r.','a,b,a*b')
ans =
As is known 4 times 5 equals 20.
\end{verbatim}
}

Instead of the formatting operator \verb!%d! we should use \verb!#r!.
In this case there is no difference between the output of \verb!sprintf! and \verb!xprintf!.

\smallskip
To demonstrate a difference between these two functions we shall consider a few more examples. 

\smallskip

{\small
\begin{verbatim}
>> a=5/7; b=3/8; 
>> sprintf('The sum of rational numbers %.3f and %.3f is %.3f.',a,b,a+b)
ans =
The sum of rational numbers 0.714 and 0.375 is 1.089.
>> a=5/7; b=3/8; 
>> xprintf('The sum of rational numbers #r and #r is #r.','a,b,a+b')
ans =
The sum of rational numbers \frac{5}{7} and \frac{3}{8} is \frac{61}{56}.
\end{verbatim}
}

\smallskip
If value of a real variable is a rational number with not very big denominator, then 
the function \verb!xprintf! recognises this and writes
value of the variable in LaTeX format as a rational number. 
As can be seen, \verb!xprintf! does not automatically augment 
a formula with dollar signs. 

\smallskip

{\small
\begin{verbatim}
>> a=sqrt(18); b=sqrt(98); 
>> xprintf('The product of irrational numbers $#r$ and $#r$ is $#r.$','a,b,a*b')
ans =
The product of irrational numbers $3\sqrt{2}$ and $7\sqrt{2}$ is $42.$
\end{verbatim}
}

\smallskip
In this example the function \verb!xprintf! recognised that the variables \verb!a! 
and \verb!b! contain square roots of integers, and appropriately wrote them in LaTeX format. 
If \verb!xprintf! fails to represent a real number in some reasonable algebraic form then it writes the number 
with four digits after the decimal point. A report about this goes to log-file. 

{\small
\begin{verbatim}
>> a=25*pi; b=7*pi; 
>> xprintf('The quotient of numbers $#r$ and $#r$ is a rational number $#r.$','a,b,a/b')
ans =
The quotient of numbers $25\pi$ and $7\pi$ is a rational number $\frac{25}{7}.$
\end{verbatim}
}

\smallskip
\noindent One can force \verb!xprintf! to write real numbers in decimal point 
form with a specified number of digits after the decimal point as follows:

{\small
\begin{verbatim}
>> a=25*pi; b=2*pi; 
>> xprintf('The quotient of numbers #7r and #3r is a rational number #5r.','a,b,a/b')
ans =
The quotient of numbers 78.5398163 and 6.283 is a rational number 12.50000.
\end{verbatim}
}

\smallskip
The operator \verb!#e! represents real numbers in floating point form.
For example,

{\small
\begin{verbatim}
>> a=25*pi; b=2*pi; 
>> xprintf('The quotient of $#e$ and $#e$ is $#e.$','a,b,a/b')
ans =
The quotient of $7.854\cdot 10^{1}$ and $6.283$ is $1.250\cdot 10^{1}$.
\end{verbatim}
}

If necessary, the number of significant digits can be given as an argument after \verb!#! as follows

{\small
\begin{verbatim}
>> a=25*pi; b=2*pi; xprintf('The quotient of $#6e$ and $#3e$ is $#4e.$','a,b,a/b')
ans =
The quotient of $7.85398\cdot 10^{1}$ and $6.28$ is $1.250\cdot 10^{1}.$
\end{verbatim}
}

\smallskip

If we wish to use in \verb!xprintf! the formatting of Matlab's \verb!sprintf! function, 
we should use the percentage symbol \verb!%! after \verb!#!. 
For example, 
{\small
\begin{verbatim}
   \verb!xprintf('What is #%5d times #%12d?','a,b')! 
\end{verbatim}
}
does the same as 
{\small
\begin{verbatim}
   \verb!sprintf('What is %5d times %12d',a,b)!.
\end{verbatim}
}
\medskip
In the following subsections we describe other \verb!xprintf! operators. 

\subsection{Coefficients {\tt c}}

The hash-tag command \verb!#c! indicates that the corresponding variable is to be treated as a coefficient. 
For example, if the corresponding variable has value $12$ then 
\verb!#cx^2! will be replaced by \verb!12x^2!.
Exceptional cases: 
   if the variable has value 1, then \verb!cx^2! will be replaced by \verb'x^2';
   if \verb!#!-variable has value $-1,$ then \verb!cx^2! will be replaced by \verb'-x^2';
   if \verb!#!-variable has value 0, then \verb!cx^2! will be replaced by empty space,
     more exactly, in this case Spike will remove all characters after \verb!#c!
     until an empty space is encountered. 

  If the \verb!#c! command has \verb!+! as a parameter, then Spike writes \verb!+! before a positive
  coefficient.

  If the value of \verb!#!-variable is not an integer then it will be written in algebraic form
  as in \verb!#r!. 

\subsection{Floating point numbers {\tt e}}
 \verb!#e!, \verb!#[+][c,s,i,f][mantissalength][,base]e!

The command \verb!#e! instructs Spike to write a real number in floating point form.
Using the optional parameter \verb![+]!, 
one can choose whether to write $+$ before a positive number or not. 
Further, the optional parameter \verb![c,s,i,f]! allows to choose between computer form (c), 
scientific form (s) or fixed point form (f), if the parameter (i) is chosen then \verb!xprintf! 
rounds the number and treats it as an integer. The other two optional parameters 
\verb![mantissalength]! and \verb![base]! are self-explanatory. 

Examples. 

{\small
\begin{verbatim}
>> [xprintf('#e','pi^4'), '  ', xprintf('#12e','exp(5)')]
ans =
9.741\cdot 10^{1}  1.48413159103\cdot 10^{2}

>> [xprintf('#"f6"e','pi^4'), '  ', xprintf('#"c5,2"e','pi^4')]
ans =
97.409091  0.11000\cdot 2^{7}
>> xprintf('#"i"e','round(pi^3)')
ans =
31
>> xprintf('#"+i0,2"e','round(pi^6)')
ans =
+1111000001
>> xprintf('#9,16e','pi^7')
ans =
B.CC4B10FA\cdot 16^{2}
\end{verbatim}
}

\bigskip

Examples. 

{\small
\begin{verbatim}
<<problemI;
  x=1000+randi(9000);
  answer('\verb!#"i0,16"e!$_{16}$',x);
@2:3;x;
  Convert the number $#r$ to the hexadecimal system.
>>

<<problemI;
  x={1001:9999}; 
  condition = (rem(x,2)~=0);
  y=x/10000;
  answer('\verb!#"f7,2"e!$_{2}$ \\',y);
@2:2;x;
Convert the number \verb!0.#r! to base 2
rounding the answer to seven digits after the binary point. 
>>
\end{verbatim}
}

\problemtype{I}
%Q1(a)
\QuesA{B}{2}\ \ \medskip (a)
  Convert the number $8361$ to the hexadecimal system.

\medskip\noindent
(A) \ \verb!162B!$_{16}$\quad
(B) \ \verb!20A9!$_{16}$\quad
(C) \ \verb!2358!$_{16}$\quad
(D) \ \verb!1CBC!$_{16}$\quad
(E) \ \verb!564!$_{16}$\quad
%Raw volume 3.36e+10

%Q1(b)
\QuesB{B}(b)
  Convert the number $5484$ to the hexadecimal system.

\medskip\noindent
(A) \ \verb!17E7!$_{16}$\quad
(B) \ \verb!156C!$_{16}$\quad
(C) \ \verb!FC3!$_{16}$\quad
(D) \ \verb!15AA!$_{16}$\quad
(E) \ \verb!1814!$_{16}$\quad

%Q1(c)
\QuesB{C}(c)
  Convert the number $6306$ to the hexadecimal system.

\medskip\noindent
(A) \ \verb!17C5!$_{16}$\quad
(B) \ \verb!C13!$_{16}$\quad
(C) \ \verb!18A2!$_{16}$\quad
(D) \ \verb!1BD8!$_{16}$\quad
(E) \ \verb!2361!$_{16}$\quad

\problemtype{I}
%Q2(a)
\QuesA{A}{2}\ \ \medskip (a)
Convert the number \verb!0.1867! to base 2
rounding the answer to seven digits after the binary point.

\medskip\noindent
(A) \ \verb!0.0011000!$_{2}$ \\\quad
(B) \ \verb!0.1000100!$_{2}$ \\\quad
(C) \ \verb!0.1110110!$_{2}$ \\\quad
(D) \ \verb!0.0010111!$_{2}$ \\\quad
(E) \ \verb!0.0111000!$_{2}$ \\\quad
%Raw volume 5.90e+19

%Q2(b)
\QuesB{D}(b)
Convert the number \verb!0.1683! to base 2
rounding the answer to seven digits after the binary point.

\medskip\noindent
(A) \ \verb!0.0100001!$_{2}$ \\\quad
(B) \ \verb!0.1111010!$_{2}$ \\\quad
(C) \ \verb!0.1010001!$_{2}$ \\\quad
(D) \ \verb!0.0010110!$_{2}$ \\\quad
(E) \ \verb!0.1100111!$_{2}$ \\\quad

\subsection{Linear combinations {\tt l}}

The command \verb!#l! writes a row of numbers as a linear combination
whose coefficients are entries of the row. 

The following examples demonstrate how to use this command. 

{\small
\begin{verbatim}
>> a=[-3 0 0 1 -8 0 -1]; xprintf('#l','a')
 ans =
 -3x_{1}+x_{4}-8x_{5}-x_{7}   % x_1, x_2, ... 
                              % are a default choice of variables of a linear combination.

>> a=[1 2 3 6]; xprintf('#"+~y"l','a')
ans =
+y_{1}+2y_{2}+3y_{3}+6y_{4}   % write y_j instead of x_j

>> a=[1 2 3 6]; xprintf('#"+~x,y,z,w"l','a')   % use x,y,z,w
                                            % instead of x_1, x2, ... 
ans =
+x_{1}+2y_{2}+3z_{3}+6w_{4}

>> a=[-1 0 3 1]; xprintf('#"~\alpha"l','a')
ans =
-\alpha_{1}+3\alpha_{3}+\alpha_{4}
\end{verbatim}
}

\medskip
Example. 

{\small
\begin{verbatim}
<<problemI;
  p={7,11};
  freeze
  A = randi(p-1,3,3);
  condition = NotZero(mod(det(A),p));
  x = randi(p-1,3,1);
  b=mod(A*x,p);
  rmdr=mod(x'*x,p);
  answer('#r',rmdr);
@3:2;p,A(1,:),b(1),A(2,:),b(2),A(3,:),b(3),p;
Solve the system of equations in~$\mbZ_{#r}$
\begin{equation*}
  \begin{split}
      #l & = #r, \\
      #l & = #r, \\
      #l & = #r, 
  \end{split}
\end{equation*}
and thereby calculate $x_1^2+x_2^2+x_3^2$ (in~$\mbZ_{#r}$).
>>
\end{verbatim}
}

\problemtype{I}
%Q2(a)
\QuesA{B}{3}\ \ \medskip (a)
Solve the system of equations in~$\mbZ_{11}$
\begin{equation*}
  \begin{split}
      4x_{1}+10x_{2}+6x_{3} & = 3, \\
      7x_{1}+7x_{2}+6x_{3} & = 0, \\
      x_{1}+7x_{2}+8x_{3} & = 8,
  \end{split}
\end{equation*}
and thereby calculate $x_1^2+x_2^2+x_3^2$ (in~$\mbZ_{11}$).

\medskip\noindent
(A) \ 6\quad
(B) \ 9\quad
(C) \ 4\quad
(D) \ 1\quad
(E) \ 7\quad
%Raw volume 2.31e+24

%Q2(b)
\QuesB{A}(b)
Solve the system of equations in~$\mbZ_{11}$
\begin{equation*}
  \begin{split}
      3x_{1}+x_{2}+x_{3} & = 7, \\
      2x_{1}+6x_{2}+9x_{3} & = 7, \\
      2x_{1}+3x_{2}+7x_{3} & = 8,
  \end{split}
\end{equation*}
and thereby calculate $x_1^2+x_2^2+x_3^2$ (in~$\mbZ_{11}$).

\medskip\noindent
(A) \ 1\quad
(B) \ 5\quad
(C) \ 6\quad
(D) \ 3\quad
(E) \ 10\quad

\bigskip
{\small
\begin{verbatim}
<<problemI;
  p={7,11,13,17};
  freeze
  A = randi(p-1,2,2);
  condition = NotZero(mod(det(A),p));
  x = randi(p-1,2,1);
  b=mod(A*x,p);
  rmdr=mod(x'*x,p);
  answer('#r',rmdr);
@3:2;p,A(1,:),b(1),A(2,:),b(2),p;
Solve the system of equations in~$\mbZ_{#r}$
\begin{equation*}
  \begin{split}
      #"~x,y"l & = #r, \\
      #"~x,y"l & = #r, 
  \end{split}
\end{equation*}
and thereby calculate $x^2+y^2$ (in~$\mbZ_{#r}$).
>>
\end{verbatim}
}

\problemtype{I}
%Q1(a)
\QuesA{A}{3}\ \ \medskip (a)
Solve the system of equations in~$\mbZ_{11}$
\begin{equation*}
  \begin{split}
      x+6y & = 3, \\
      9x+2y & = 7,
  \end{split}
\end{equation*}
and thereby calculate $x^2+y^2$ (in~$\mbZ_{11}$).

\medskip\noindent
(A) \ 10\quad
(B) \ 7\quad
(C) \ 6\quad
(D) \ 4\quad
(E) \ 1\quad
%Raw volume 4.72e+27

%Q1(b)
\QuesB{C}(b)
Solve the system of equations in~$\mbZ_{13}$
\begin{equation*}
  \begin{split}
      11x+5y & = 9, \\
      10x+12y & = 8,
  \end{split}
\end{equation*}
and thereby calculate $x^2+y^2$ (in~$\mbZ_{13}$).

\medskip\noindent
(A) \ 4\quad
(B) \ 2\quad
(C) \ 0\quad
(D) \ 5\quad
(E) \ 8\quad

\subsection{Matrices {\tt m}}

 \verb!#m   #[]m   #()m or #Nm, #[]Nm, #()Nm,! where $N$ is a positive integer.

 This command writes a matrix \verb!#!-variable
 in LaTeX format. The command \verb!#m! writes a matrix without brackets,
 the command \verb!#[]m! writes a matrix with square brackets,
 the command \verb!#()m! writes a matrix with round brackets.

 Entries of the matrix \verb!#!-variable  
 can be non-integer rational numbers, in which case Spike 
 will write entries as fractions in LaTeX format. 
 If there is a parameter \verb!N! in \verb!#m! command then Spike will write entries of the 
 matrix as fixed point real numbers with N digits after the decimal point.

\medskip
Examples. 

{\small
\begin{verbatim}
<<problemI;
  A = randi(14,4,4)-7;
  d = round(det(A));
  answer('$#r$',d);
@1;A;
Find determinant of the matrix 
$$#[]m.$$
>>
\end{verbatim}
}

\problemtype{I}
\QuesA{B}{1}
Find determinant of the matrix
$$\left[
  \begin{matrix}
    7&-4&-4&7\\
    6&6&7&5\\
    -4&-1&-3&-4\\
    0&4&-6&6\\
  \end{matrix}
\right].$$

\medskip\noindent
(A) \ $-147$ \quad 
(B) \ $-940$ \quad 
(C) \ $342$ \quad 
(D) \ $297$ \quad 
(E) \ $-379$ \quad 

\medskip

{\small 
\begin{verbatim}
<<problemI;
  A = randi(14,2,2)-7;
  d = round(det(A));
  condition=(d~=0);
  answer('$#()m$',inv(A));
@2::3;A;
Find inverse of the matrix 
$$
  #()m.
$$
>>
\end{verbatim}
}

\problemtype{I}
\QuesA{C}{2}
Find inverse of the matrix
$$
\left(
  \begin{matrix}
    3&4\\
    0&-5\\
  \end{matrix}
\right).
$$

\medskip\noindent
(A) \ $\left(
  \begin{matrix}
    \frac{1}{13}&-\frac{3}{26}\\
    \frac{1}{13}&\frac{7}{52}\\
  \end{matrix}
\right)$
\quad 
(B) \ $\left(
  \begin{matrix}
    1&\frac{3}{4}\\
    0&-\frac{1}{4}\\
  \end{matrix}
\right)$
\quad 
(C) \ $\left(
  \begin{matrix}
    \frac{1}{3}&\frac{4}{15}\\
    0&-\frac{1}{5}\\
  \end{matrix}
\right)$

\subsection{Polynomials {\tt p, <p, >p}}

 This command writes a polynomial in LaTeX format
 with coefficients taken from a vector \verb!#!-variable. 
 The command \verb!#<p! (respectively, \verb!#>p!) tells Spike to write the polynomial
 in the order of increasing (respectively, decreasing) powers.
 If the order is not specified, then Spike uses the default order 
 of powers, which is \verb!#<p!. 

 For example, if the \verb!#!-variable \verb!a! has value \verb![1,0,0,0.5,0,0.1]!
 then \verb'#p' will be replaced by 
\begin{verbatim}
  1+\frac{1}{2}x^{3}+\frac{1}{10}x^{5},
\end{verbatim}
which gives
$$
  1+\frac{1}{2}x^{3}+\frac{1}{10}x^{5}.
$$
For \verb!a! with the same value \verb'#>p' will be replaced by 
$$
  \frac{1}{10}x^{5}+\frac{1}{2}x^{3}+1.
$$
A term with zero coefficient is dropped. 
Spike also drops coefficient of a term if it is +1 or -1,
only the sign is written.

\medskip
Example.

{\small
\begin{verbatim}
<<problemI;
  p={7,11,13};
  freeze
  N={8:19}; 
  a=zeros(1,N);
  a(1:3)=1+randi(p-2,1,3); 
  a(N)=1;
  c=1+randi(p-2);
  answ=PolyEval(a,c,p);
  answer('$#r$',answ);
@2:2;a,p-c,p;
Let $f(x) = #>p$ and $g(x) = x+#r$ be two polynomials from~$\mbZ_{#r}[x].$
Find remainder when $f(x)$ is divided by $g(x).$ 
>>
\end{verbatim}
}

\problemtype{I}
%Q1(a)
\QuesA{E}{2}\ \ \medskip (a)
Let $f(x) = x^{7}+2x^{2}+12x+5$ and $g(x) = x+8$ be two polynomials from~$\mbZ_{13}[x].$
Find remainder when $f(x)$ is divided by $g(x).$

\medskip\noindent
(A) \ $12$\quad
(B) \ $5$\quad
(C) \ $9$\quad
(D) \ $2$\quad
(E) \ $6$\quad
%Raw volume 4.05e+22

%Q1(b)
\QuesB{E}(b)
Let $f(x) = x^{12}+5x^{2}+9x+10$ and $g(x) = x+2$ be two polynomials from~$\mbZ_{11}[x].$
Find remainder when $f(x)$ is divided by $g(x).$

\medskip\noindent
(A) \ $7$\quad
(B) \ $8$\quad
(C) \ $9$\quad
(D) \ $2$\quad
(E) \ $5$\quad

\subsection{Permutations {\tt 0q  1q  2q}}
We shall identify an $n$-permutation given by standard $2\times n$ table
with the second row of this table. 

The command \verb!#0q! writes a permutation in its standard form as a table.
The command \verb!#1q! writes a permutation as a product of cycles including 
trivial length one cycles. 
The command \verb!#2q! writes a permutation as a product of cycles omitting
trivial length one cycles. 

\medskip
Example. 
{\small
\begin{verbatim}
<<problemI;
  n=7;
  freeze
  a = randperm(n);
  answer('#1q',a);
@2;a;
Write the permutation
$$
  #0q
$$
as a product of disjoint cycles.
>>
\end{verbatim}
}

\QuesA{C}{2}
Write the permutation
$$
\left(
  \begin{array}{ccccccc}
    1&2&3&4&5&6&7\\
    3&1&7&6&4&5&2
  \end{array}
\right)
$$
as a product of disjoint cycles.

\medskip\noindent
(A) \ (175)(23)(46) \quad 
(B) \ (13)(26745) \quad 
(C) \ (1372)(465) \quad 
(D) \ (127536)(4) \quad 
(E) \ (1564237) \quad 

\subsection{Algebraically representable numbers {\tt r, +r, +Nr}}

This command tries to write a real number as an algebraic expression, 
for example, \verb!1/3! will be written as \verb!\frac{1}{3}!, not as \verb!0.3333!. 
If it fails to do this, then it will write the real number with four significant digits. 

The option \verb!#+r! is a signed version of \verb!#r!.

Examples. 
{\small
\begin{verbatim}
>> a=sqrt(125); xprintf('#r','a')
ans =
5\sqrt{5}
>> b=4/(40*pi); xprintf('#r','b')
ans =
\frac{1}{10\pi}
>> c=log(50)-log(2); xprintf('#r','c')
ans =
\ln\,25


<<problemI;
  p,r,s!={2,3,5,7,11};
  x=p^2*r*s;
  answer('$#r\sqrt{#r}$',p,r*s);
@2:2:6; x;
Simplify $\sqrt{#r}.$
>>
\end{verbatim}
}

\problemtype{I}
%Q1(a)
\QuesA{F}{2}\ \ \medskip (a)
Simplify $\sqrt{315}.$

\medskip\noindent
(A) \ $2\sqrt{15}$\quad
(B) \ $5\sqrt{21}$\quad
(C) \ $5\sqrt{77}$\quad
(D) \ $11\sqrt{35}$\quad
(E) \ $11\sqrt{21}$\quad
(F) \ $3\sqrt{35}$\quad
%Raw volume 60

%Q1(b)
\QuesB{F}(b)
Simplify $\sqrt{308}.$

\medskip\noindent
(A) \ $11\sqrt{10}$\quad
(B) \ $2\sqrt{33}$\quad
(C) \ $5\sqrt{77}$\quad
(D) \ $3\sqrt{22}$\quad
(E) \ $5\sqrt{33}$\quad
(F) \ $2\sqrt{77}$\quad

\bigskip

{\small
\begin{verbatim}
<<problemI;
  p,r,s!={2,3,5,7,11};
  freeze 
  a,b!={1:20};
  condition=(gcd(a,p*r)==1);
  condition=(gcd(b,p*s)==1);
  x=a/(p*r);
  condition=(x<1);
  y=b/(p*s);
  condition=(y<1);
  answer('$#r$',x+y);
@2:2:6; x,y;
Calculate $#r+#r.$
>>
\end{verbatim}
}

\problemtype{I}
%Q2(a)
\QuesA{C}{2}\ \ \medskip (a)
Calculate $\frac{16}{55}+\frac{8}{15}.$

\medskip\noindent
(A) \ $\frac{194}{165}$\quad
(B) \ $\frac{17}{165}$\quad
(C) \ $\frac{136}{165}$\quad
(D) \ $\frac{73}{165}$\quad
(E) \ $\frac{41}{33}$\quad
(F) \ $\frac{16}{33}$\quad
%Raw volume 8.66e+06

%Q2(b)
\QuesB{B}(b)
Calculate $\frac{7}{22}+\frac{1}{6}.$

\medskip\noindent
(A) \ $\frac{31}{33}$\quad
(B) \ $\frac{16}{33}$\quad
(C) \ $\frac{32}{33}$\quad
(D) \ $\frac{41}{33}$\quad
(E) \ $\frac{13}{33}$\quad
(F) \ $\frac{50}{33}$\quad

\subsection{Strings {\tt s}}

 This command works exactly as Matlab's \verb!%s! command.
 
\subsection{Lines of tables {\tt  t}}

This command returns a string which contains elements of a vector
separated by a character \verb'&'. For example, if  \verb!a=[3,5,7]! then \verb'#&t'
will be replaced by \verb'3&5&7'.

In fact, one can use any other non-letter character instead of \verb'&', for example, \verb','.
For instance, \verb'#,t' will be replaced by \verb'3,5,7'.
 
If elements of a vector are real numbers, then after the separator symbol one 
can give the number of digits to be written.  For example, \verb!#&~2t! will 
write elements of the vector with two digits after the decimal point and these 
numbers will be separated by \verb!&!.

\medskip
Example. 

{\small
\begin{verbatim}
<<problemI;
  a=randi(20,1,7)-10;
  b=randi(20,1,7)-10;
  answ=a*b';
  answer('$#r$',answ);
@1;a, b;
Find the scalar product of vectors $\mathbf a = (#,t)$ 
and $\mathbf b = (#,t).$
>>
\end{verbatim}
}

\QuesA{C}{1}
Find the scalar product of vectors $\mathbf a = (8,-3,4,-8,-4,-9,-7)$ and
$\mathbf b = (-6,2,2,-5,6,2,8).$

\medskip\noindent
(A) \ $13$\quad 
(B) \ $-79$\quad 
(C) \ $-104$\quad 
(D) \ $-164$\quad 
(E) \ $-103$\quad 

\medskip
Here the components of vectors are separated by commas, since we gave the comma as a parameter to the hash-tag 
operator \verb!#t!. 
The comma in the hash-tag operator \verb!#,t! tells \Spike\ to separate components
of the vector by commas. Instead of commas one can use any non-letter character, for example,
\ \verb!#:t!, \ \verb!#;t! etc. 
Otherwise, the separator should be taken in double quotes, such as, for example, \verb!#",\quad"t!,
in case we wish to insert some space between components of a vector. 
The operator \verb!#t! does not augment a vector with brackets.

\subsection{Powers {\tt w}}

 This command writes an integer \verb!#!-variable, say, n, as a power.
 For example, if \verb!n=-3!, then \verb' a#w' will be replaced by \verb' a^{-3}'.
 Exceptional cases: 
     If \verb!n = 1!, then \verb' a#w' will be replaced by \verb'a'.
     If \verb!n = 0!, then \verb' a#w' will be replaced by empty string;
     more exactly, in this case Spike will remove all characters 
     preceding \verb!#w! until an empty space is encountered.
       For example, \verb'AB \mathbf{C}#w' will be replaced by \verb'AB ',
       if the value of the corresponding \verb!#!-variable is zero.

If the \verb!#!-variable has a rational non-integer value then Spike will 
write the power as a fraction. For example, if \verb!#!-variable 
\verb!n! is \verb!3/7!, then \verb' a#w' will be replaced by \verb' a^{\frac{3}{7}}'.

As an example we will randomise this question:

{\it Simplify the expression
$$
  \frac{(a^{-5}b^{4})^{-4}(a^{-6}c^{9})^{7}}{(b^{6}c^{8})^{-1}(a^{-7}b^{-6})^{-1}}.
$$
}

{\small
\begin{verbatim}
<<problemI;
  x,y,z!<{'a':'n','p':'z'};
  freeze
  n1,n2,n3!={-3:-1,1:5};
  m1,m2,m3!={-3:-1,1:5};
  k1,k2,k3!={-3:-1,1:5};
  l1,l2,l3!={-3:-1,1:5};
  p1=n1*n2+m1*m3-l1*l3;
  p2=n2*n3-k1*k3-l2*l3;
  p3=m2*m3-k2*k3;
  answ1=xprintf('#s#w #s#w #s#w','x,p1,y,p2,z,p3');
  answer('$#s$',answ1);
@1:2;x,n1,y,n2,n3,x,m1,z,m2,m3,y,k1,z,k2,k3,x,l1,y,l2,l3;
Simplify the expression
$$
  \frac{(#s#w#s#w)^{#r} (#s#w#s#w)^{#r}} {(#s#w#s#w)^{#r} (#s#w#s#w)^{#r}}.  
$$
>>
\end{verbatim}
}

\problemtype{I}
\QuesA{C}{1} (a)
Simplify the expression
$$
  \frac{(hk^{3})^{-1} (h^{4}l^{-3})^{3}} {(k^{2}l^{5})^{-2} (h^{-2}k^{2})^{-3}}.
$$

\medskip\noindent
(A) \ $h^{14} k^{5} l^{-8}$ \quad 
(B) \ $h^{2} k^{-1} l^{-6}$ \quad 
(C) \ $h^{9} k^{7} l$ \quad 
(D) \ $h^{-25} k^{2} l^{-11}$ \quad 
(E) \ $h^{25} k^{-27} l^{-11}$ \quad 

\QuesB{B} (b)
Simplify the expression
$$
  \frac{(d^{3}e^{-2})^{-3} (d^{2}t^{-1})^{-2}} {(e^{5}t^{2})^{-2} (d^{-2}e^{3})^{-3}}.
$$

\medskip\noindent
(A) \ $d^{3} e^{3} t^{-13}$ \quad 
(B) \ $d^{-16} e^{25} t^{6}$ \quad 
(C) \ $ e^{6} t^{-9}$ \quad 
(D) \ $d e^{-1} t^{-10}$ \quad 
(E) \ $d^{9} e^{-24} t^{-4}$ \quad 

\subsection{ODE's {\tt y}}

Examples.

{\small
\begin{verbatim}
>>a=[1,1,2];
>> xprintf('#y','a')
ans =
      y\ddash+y\dash+2y
>> t=[1,4,-3,4,0,1];
>> xprintf('#y','t')
ans =
      y^{(5)}+4y^{(4)}-3y\dddash+4y\ddash+y
\end{verbatim}
}
Here \verb!y\ddash! is a LaTeX command defined to produce \verb!y''!. This is done to avoid a possible conflict.

\subsection{Categorical variables {\tt b}}

If a categorical variable $x$ has $n$ values \verb!value1!, \ldots, \verb!valuen!
then 
\begin{verbatim}
   xprintf('#"value1~value2~...~valuen"b','x')
\end{verbatim}
will return \verb!valuek! if $x=k-1.$

Examples.

{\small
\begin{verbatim}
>> xprintf('#"No~Yes"b','1')
ans = Yes

>> y=2; xprintf('#"Elliptic~Parabolic~Hyperbolic"b','y')
ans = Hyperbolic
\end{verbatim}
}

\subsection{Variation of parameters of a hashtag command}
Sometimes we may need to variate a parameter of a hashtag command.

\medskip 
Example. 

{\small
\begin{verbatim}
<<problemI;
  x=1000+randi(9000);
  answer('#r',x);
@1:2;x;
  Convert the number \verb!#"i0,3"e! $_{3}$ given in base $3$ to the decimal system.
>>
\end{verbatim}
}

If we wish to variate the base $3$ we can do the following:

{\small
\begin{verbatim}
<<problemI;
  b={3:7};
  freeze
  x=1000+randi(9000);
  answer('#r',x);
@1:2;b,b,x;
  Convert the number \verb!##"i0,#r"e! $_{#r}$ to the decimal system.
>>
\end{verbatim}
}

Here we have a \emph{double-hashtag} operator \verb!##"i0,#r"e!. 
They work in the same way as usual hashtag operators,
but with one exception: they are executed after all hashtag operators. 
In the example above Spike first executes the two commands \verb!#r! and only after that 
the command \verb!##"i0,#r"e!, in which by the time of its execution
the hashtag operator \verb!#r! has been replaced by a number from the range $3..7.$ 
As the example above shows, in the at-line the variables of hashtag operators (\verb!b,b! in the example) 
should precede the variables of double-hashtag operators (\verb!x! in the example).

\section{Marking}
When \Spike\ processes an spk-file it appends to the output \LaTeX-file
the following information about assignment questions:
\begin{enumerate}
  \item type of the question-block (A,B,D,I,R,T,G, H or V) which generated the question,
  \item the number of multiple choice answers
  \item question number,
  \item weight of question,
  \item marking specifications,
  \item the correct answer to each randomised version of the question.
\end{enumerate}
This information is written in strings which start respectively with
the following identifiers: 
\begin{verbatim}
PrbType: 
NAltAns: 
Questns: 
Weights: 
Specfcs: 
\end{verbatim}

All this information is written in the output \LaTeX-file after the \verb!\end{document}! command,
so \LaTeX\ ignores it. 
This information is used for marking purposes. Everyone can decide how to process this information,
but here I present my own solution of this task.

\smallskip
For example, assume that in a topic with topic code \verb!MATH3731! there are sixteen
students with student numbers 
{\small
\begin{verbatim}
2198080
2189173
... 
2155986
\end{verbatim}
}
\noindent 
These numbers are kept in the file \verb!StudList.txt! in \verb!MATH3731\! folder.
Upon running the command \verb!>>spike('MATH3731',3,1)!, 
\Spike\ generates 16 assignments, one for each student.

At the end of the output assignment \LaTeX-file for the third assignment after the command
\verb!\end{document}! \Spike\ inserts the following information:

{\small
\begin{verbatim}
Topic code:          MATH3731
Assignment number:   3
Version:             1
Number of questions: 22
Types of questions: 4B 11I 1H 6T 


%MARKING
PrbType: T  I  I  I  I |I  I  T  B   I |T  T  I  I  I |B   B  B  I  T |T  H  
NAltAns: 5  5  5  5  5 |5  5  5  2   5 |5  5  5  5  5 |2   2  2  5  5 |5  5  
Questns: 1  2  3  4  5 |6  7  8  9   0 |1  2  3  4  5 |6   7  8  9  0 |1  2  
Weights: 3  1  2  1  1 |2  3  3  2   3 |2  2  1  2  2 |2   2  2  3  3 |3  2  
Specfcs: .  .  .  .  . |.  .  .  .   . |.  .  .  .  . |.   .  .  .  . |.  .  
2198080: 12 AD BB DB EA|ED CC 12 ABB DB|12 12 CC AD DE|AAA AA BA CA 02|13 11 
2189173: 22 EE AC CE DE|AC DE 22 AAA CE|22 22 CB DB DB|AAA AB BB EB 12|21 22 
2164183: 33 AA CE DA EE|CC CB 12 BBB BA|20 01 DC EB BA|AAB AB AB DD 31|22 11 
2197223: 34 EB CD CA DD|DE DE 21 BAB EE|21 11 BE EE AD|BBB BA AB DE 22|31 11 
2197268: 22 DC DA AC DE|EB EA 12 ABB DB|12 01 ED DA EB|BAB BB AA EC 31|11 32 
2172358: 22 BC BA EE DC|AB CA 03 BBB BC|31 21 DB DA ED|BBA AB AA DD 21|20 22 
2176365: 23 DB AB BC AC|EA CA 22 ABB DB|21 32 DE CA AB|BAB BA BB BB 21|11 11 
2177605: 32 CD BD AB EE|AB CD 23 BAA EB|31 42 BD AD AE|AAB BB BB AB 22|31 32 
2176675: 22 DE CA CE CC|EC BA 22 BBA AB|31 42 DA BD AC|BAB BB BB AC 22|21 32 
2148736: 11 CC BB AA DD|CE DA 32 AAB AA|23 21 DC CC DE|ABA AB AB AC 12|22 32 
2158744: 11 BE BD EB ED|AD BC 12 ABA EC|22 21 EA BC AE|AAA AA AA AA 21|22 32 
2196837: 24 CA CE BB DE|BA AD 12 BBB DE|32 21 DD DA EE|BBA AA BB BB 12|12 20 
2196839: 23 AC EB EC EC|DA AD 22 BBB AD|21 11 EC DD DA|ABB BA BB DA 21|22 13 
2155986: 34 DA AB ED BB|BE EA 22 ABA CB|32 02 BE BE BA|BAB AA AA CE 22|22 02 
-------

\end{verbatim}
}

\smallskip
Students ID's don't need to be sorted.
The above information is used by a marking program \verb!spikemark! to mark the assignment.
The function \verb!spikemark! has four arguments: topic code, assignment number,
a number showing how much the assignment is worth in percents of total assessment,
and an argument \verb!Sort! which can take two values 0 or 1. If \verb!Sort! is zero,
then the answer strings are not sorted, if \verb!Sort! is one, then
the answer strings are sorted with respect to the last three digits of students ID's,
while the first four digits are replaced by stars.

The command 
\begin{verbatim}
>> spikemark('MATH3731',3,1,7,1)
\end{verbatim}
marks an assignment; the number 7 indicates the maximum mark for it. 
This command should be run from the folder \verb!Spike\!, and it assumes that the assignment file
\verb!A3V1.tex! is kept in the folder \verb!\Spike\MATH3731\Final!.

The function \verb!spikemark! scans the assignment file looking for a line which starts with
\verb!%MARKING!, since \verb!spikemark! knows that marking information is given immediately after this line.
After \verb!%MARKING! line \verb!spikemark! expects to find the lines
\verb!PrbType!, \verb!NAltAns!, etc, in this order.
After these lines \verb!spikemark! expects to find correct answers to assignment questions
preceded by students' IDs, as in the example shown above until it encounters the ``seven dash line''.
Students' answers \verb!spikemark! takes from an email submissions file, which in this case must have the name
\verb!MATH3731\Submissions\A3V1.txt!. 
The answers in this file should be presented in the following form (the names are made-up):
{\small
\begin{verbatim}
Ray Smith MATH3731 A3V1 2172358 22 BC BA EE DC, AB CA 12 BBB BC, 21 21 ...! Wed 10:59 PM 17 KB		
Jay Fox MATH3731 A3V1 2197268 32 DC DA AC AE, EB EA 31 ABB DB, 00 12 ...! Wed 10:49 PM	17 KB		
Amy Hunt MATH3731 A3V1 2189173 33 EE AC CE DE, AC DE 33 AAA BE, 22 11 ...! Wed 9:59 PM	17 KB		
...
\end{verbatim}
}
I get this information by copying the content of a Microsoft Outlook email folder, which collects student
email submissions, into the submissions folder. Namely, the Outlook's \verb!Inbox! folder has the subfolder 
\verb!Assignment submissions\MATH3731! and this subfolder has eight subfolders \verb!MATHA1V1!, 
\verb!MATHA2V1!, \ldots, \verb!MATHA4V2!. 
The function \verb!spikemark! scans each line of the submissions file \verb!A3V1_2017.txt! for the string 
\verb!MATH3731 A3V1! after which it expects to find a seven-digit student ID and a list 
of the students' answers followed by \verb'!'.
Everything before \verb!MATH3731 A3V1! and after \verb'!' the function \verb!spikemark! ignores. 

\medskip
The command \verb!spikemark('MATH3731',3,1,7,1)! prepares a LaTeX file \verb!MATH3731A3V1_marks.tex!,
which is saved in the subfolder \verb!\Spike\MATH3731\Marks!. 
This file contains the processed marking information.
The following table is produced by \verb!spikemark('MATH3731',3,1,7,1)!.

\newpage
\begin{center}
Flinders University 2017 Semester 2 \\
MATH3731 Assignment 3 (version 1) marks as by 23-Oct-2017
\end{center}

\begin{center}
\begin{tabular}{|r|l|c|r|r|}\hline\hline
&{\small Stud.\,ID}&Answers  &{\small p-ts}&{\small mark}\\
\hline\hline
 1&\verb!****080!&\verb!12 ++ ++ ++ ++|++ ++ ++ +++ D+|++ 12 ++ ++ ++|AAA ++ ++ ++ 0+|+3 ++ !&31& 4.6\\ \hline %2198080  Abhishek Kumar     
 2&\verb!****173!&\verb!22 ++ ++ ++ ++|++ ++ 22 +++ C+|++ 22 ++ ++ ++|+++ ++ ++ +B ++|++ 22 !&31& 4.6\\ \hline %2189173  Rehab Alrashidi    
 3&\verb!****183!&\verb!33 ++ ++ ++ ++|C+ ++ 1+ +++ ++|+0 ++ ++ +B B+|AA+ ++ +B D+ 31|+2 ++ !&20& 3.0\\ \hline %2164183  Ebraheem Shaibah   
 4&\verb!****223!&\verb!+4 ++ ++ ++ ++|++ ++ ++ +++ ++|++ 11 ++ ++ ++|BB+ +A ++ ++ +2|++ ++ !&35& 5.2\\ \hline %2197223  Yau Fuk Wong       
 5&\verb!****268!&\verb!2+ ++ ++ ++ D+|++ ++ 12 +++ ++|12 01 ++ ++ ++|BAB ++ ++ E+ 31|+1 3+ !&23& 3.4\\ \hline %2197268  Sukhdeep Kaur      
 6&\verb!****358!&\verb!++ ++ ++ ++ ++|++ ++ 03 +++ ++|3+ ++ ++ ++ ++|+++ +B AA ++ ++|+0 ++ !&35& 5.2\\ \hline %2172358  Zayed Asiri        
 7&\verb!****365!&\verb!2+ DB ++ ++ ++|++ ++ ++ +++ ++|+1 +2 ++ +A ++|BA. ++ BB ++ 21|++ ++ !&30& 4.5\\ \hline %2176365  Kirandeep Kaur     
 8&\verb!****605!&\verb!++ ++ ++ ++ ++|++ ++ ++ +++ ++|+1 4+ ++ ++ ++|+++ ++ ++ A+ ++|3+ ++ !&37& 5.5\\ \hline %2177605  Pinkham Thanavanh  
 9&\verb!****675!&\verb!22 ++ ++ ++ ++|++ B+ 2+ +++ ++|++ ++ ++ +D ++|+++ ++ ++ +C 22|++ ++ !&30& 4.5\\ \hline %2176675  Li Ying            
10&\verb!****736!&\verb!11 ++ ++ ++ ++|++ ++ ++ +++ ++|++ +1 ++ ++ ++|+++ ++ ++ ++ +2|+2 ++ !&36& 5.4\\ \hline %2148736  Mohammed Balkhudher
11&\verb!****744!&\verb!++ ++ ++ ++ ++|++ ++ ++ +++ ++|++ ++ ++ ++ A+|+++ ++ ++ ++ ++|++ ++ !&45& 6.7\\ \hline %2158744  Eman Alamri        
12&\verb!****837!&\verb!.. .. .. .. ..|.. .. .. ... ..|.. .. .. .. ..|... .. .. .. ..|.. .. !& 0& 0.0\\ \hline %............................
13&\verb!****839!&\verb!++ ++ ++ ++ ++|++ ++ +2 +++ AD|+1 +1 ++ ++ ++|AB. ++ ++ D+ 21|++ ++ !&29& 4.3\\ \hline %2196839  Money Garg         
14&\verb!****986!&\verb!++ ++ ++ ++ ++|++ ++ ++ +++ ++|++ 02 ++ ++ ++|+++ ++ ++ ++ 2+|2+ ++ !&39& 5.8\\ \hline %2155986  Mohammed Alammar   
\hline
         &Questns&\verb!1  2  3  4  5 |6  7  8  9   0 |1  2  3  4  5 |6   7  8  9  0 |1  2  ! &       &\\\hline\hline
         &Repeats&\verb!ab ab ab ab ab|ab ab ab abc ab|ab ab ab ab ab|abc ab ab ab ab|ab ab ! &max&max\\\hline\hline
         &Weights&\verb!3  1  2  1  1 |2  3  3  2   3 |2  2  1  2  2 |2   2  2  3  3 |3  2  ! &47&7\\\hline\hline
\end{tabular}
\end{center}

\smallskip\smallskip
If a student's answer is incorrect, then the correct 
answer \verb!A,B,C,...! or \verb!"-"! is shown in the table.

\smallskip \verb!"-"! means that student's answer 
is incorrect and the correct answer will be given later.

\smallskip \verb!"+"! means that student's answer is correct.

\smallskip \verb!"."! means that the answer was not given.

%\smallskip If an error message is shown instead of an answer line\\
%you should resubmit your corrected answer line again ASAP.

\bigskip
\noindent The number of students enrolled in this topic: 14.\\
\noindent The number of submissions: 13.\\
\noindent The average mark for this assignment:  4.823.\\
\noindent The standard deviation of non-zero marks:  0.978.

\bigskip\noindent The rounded numbers of correct answers given to
 each question averaged over the number of repeats: 

{\small \begin{verbatim}
      1   2   3   4   5   6   7   8   9  10  11  12  13  14  15  16  17  18  19  20  21  22
      7  12  13  13  13  13  13   9  13  11  10   6  13  12  12   8  12  11  10   6  10  12
%:   50  92 100 100  96  96  96  65 100  85  73  46 100  88  92  59  92  81  77  46  73  88
     **                          --              **              --              **        

\end{verbatim}
}

\bigskip

\noindent Mark distribution:

\smallskip 

\begin{verbatim}
   0.0- 0.4:  *
   0.5- 1.4:  
   1.5- 2.4:  
   2.5- 3.4:  **
   3.5- 4.4:  ***
   4.5- 5.4:  *****
   5.5- 6.4:  **
   6.5- 7.0:  *
\end{verbatim}

\smallskip \smallskip \smallskip
\hrule

\bigskip
The \verb!Specfcs! line consists of four characters: \verb!".+0n"!.
The dot character \verb!"."! means that the question should be marked as usual.
The plus character \verb!"+"! means that all students' answers to this question should be
considered as correct.
The zero character \verb!"0"! means that the question should be ignored, that is,
student's answers should be ignored if there are ones, whether they are correct or not,
and the question's weights should be ignored, that is, to be assumed to be equal to zero.
Finally, the character \verb!n! means that the correct answer to
this question should not be given to students.

\end{document}